\documentclass[aps,prd,floatfix,showpacs,showkeys,superscriptadress,amsmath,unsortedaddress,nofootinbib,twocolumn]{revtex4-1}
\usepackage{mathtools,amsmath,amssymb,bbm,bm,slashed}
\usepackage{graphicx,float}
\usepackage{hyperref}
\usepackage{color,bbold}
\hypersetup{colorlinks=true,linkcolor=blue,citecolor=blue,urlcolor=blue}
\usepackage{morefloats,subfigure}
\newcommand{\sF}{\scriptscriptstyle{f}}
\newcommand{\sB}{\scriptscriptstyle{B}}
\newcommand{\sperp}{\scriptscriptstyle{\perp}}
\newcommand{\shp}{\shortparallel}
\newcommand{\be}{\begin{eqnarray}}
\newcommand{\ee}{\end{eqnarray}}
\newcommand{\nn}{\nonumber \\}
\newcommand{\wt}[1]{\widetilde#1}

\DeclareMathOperator{\sech}{sech}
\DeclareMathOperator{\csch}{csch}
\def\({\left(}
\def\){\right)}
\allowdisplaybreaks
\begin{document}
\title{Neutral pion mass in the linear sigma model coupled to quarks at arbitrary magnetic field}
\author{Aritra Das,}
\email{aritra.das@saha.ac.in} 
\affiliation{HENPP Division, Saha Institute of Nuclear Physics, HBNI, 
	1/AF Bidhan Nagar, Kolkata 700064, India}
\author{Najmul Haque,}
\email[]{nhaque@niser.ac.in}
\affiliation{School of Physical Sciences, National Institute of Science Education and Research,\\ HBNI, Jatni 752050, India}

\begin{abstract}
	We calculate the neutral pion mass in the presence of an external magnetic field of  arbitrary strength in the framework of the linear sigma model coupled to quarks at zero temperature. We find nonmonotonic behavior of the pion mass as a function of magnetic field. We are also able to reproduce existing results for the weak-field approximation. 
\end{abstract}

\maketitle

\section{Introduction}
In heavy-ion collision (HIC) experiments, a very strong anisotropic magnetic field ($\sim 10^{19}$ G) is generated in peripheral collisions due to the relative motion of the colliding ions~\cite{Skokov:2009qp,Zhong:2014cda,Tuchin:2010gx,Tuchin:2014iua,Tuchin:2013bda}. The direction of the generated magnetic field is perpendicular to the reaction plane. Apart from HIC experiments, finite magnetic fields also involved in the interior of dense astrophysical objects like compact stars, magnetars~\cite{Duncan:1992hi}, and in the early Universe. The effects of such magnetic fields on fundamental particles cannot be neglected and a detailed understanding of their effects on elementary particles is essential. 

One such effect is the behavior of meson masses as a function of the strength of the magnetic field. The study of magnetic -field-dependent meson masses is the subject of active research. In Ref.~\cite{Bali:2017ian}, the authors studied the masses of light mesons -- namely, the charged and neutral pions $(\pi^{\pm},\pi^0)$ and rho mesons $(\rho^{\pm},\rho^0)$ in the presence of an external electromagnetic field in the framework of lattice quantum chromodynamics (LQCD) and showed that the magnetic-field-dependent neutral pion mass decreases with the magnetic field strength.

 Apart from the LQCD calculations, various effective QCD models have been used to study the properties of meson masses in the presence of the magnetic field. These include chiral-perturbation theory~\cite{Agasian:2001ym,Andersen:2012dz}, pseudoscalar and pseudovector pion-nucleon interaction models~\cite{Adhya:2016ydf,Mukherjee:2017dls}, the Nambu-Jona-Laisino(NJL) model and its extension~\cite{Fayazbakhsh:2012vr,Coppola:2018vkw,Avancini:2015ady,GomezDumm:2017jij,Zhang:2016qrl,Avancini:2018svs,Avancini:2016fgq}, the Polyakov-loop extension of the NJL (PNJL) model~\cite{Gatto:2010qs,Kashiwa:2011js}, the Polyakov loop quark-meson model~\cite{Mizher:2010zb,Skokov:2011ib}, and quark-meson model~\cite{Fraga:2008qn,Frasca:2011zn,Rabhi:2011mj,Andersen:2011ip,Andersen:2012bq}.
 
 The linear sigma model (LSM) is one of the oldest and simplest model in pre-QCD era, and was originally proposed by Gell-Mann and L{\'e}vy~\cite{GellMann:1960np} to describe pion-nucleon interactions. Many global symmetries of QCD are exhibited in the LSM. Later, this simple model was also used to study chiral phase transitions~\cite{Petropoulos:2004bt}, magnetic and thermomagnetic corrections to the $\pi$-$\pi$ scattering length~\cite{Loewe:2017kiw,Loewe:2019xtn}, pion condensates in the presence of a magnetic field~\cite{Loewe:2013coa}, and many more.
 
 The addition of light quarks to the LSM Lagrangian density has given a new dimension to the existing model. It is called the linear sigma model coupled to quarks (LSMq). The study of QCD phase diagrams has been carried out within the framework of the LSMq in Refs.~\cite{Ayala:2019skg,Ayala:2014jla,Ayala:2017ucc}. In Ref.~\cite{Scavenius:2000qd}, the authors not only studied the thermodynamics of the QCD phase diagram and chiral transition, but also compared it with that obtained from the NJL model. Also, the LSMq was recently used to study, the magnetized QCD phase diagram~\cite{Ayala:2015lta} and inverse magnetic catalysis~\cite{Ayala:2014gwa}.     
  
In Ref.~\cite{Ayala:2018zat} the authors studied the magnetic-field-dependent neutral pion mass within the weak magnetic field approximation. In this article, we generalize the calculation which is valid at any value of the magnetic field strength. 

The paper is organized as follows. In Sec.~\ref{lsm}, we review the LSMq with an explicit symmetry-breaking term to account for the nonzero pion mass. In Sec.~\ref{sec:pi0} we compute the  magnetic field correction of the neutral pion self-energy at one-loop order that comprises the quark-antiquark contribution (Sec.~\ref{sec:Piff}), the charged meson contribution (Sec.~\ref{sec:Pipm}). In Sec.~\ref{sec:Pi_mass} we compute the neutral pion mass by solving the pion dispersion relation, which is the sum of the quark-antiquark and meson contributions. In Sec.~\ref{sec:asymp} the pion mass in the weak-field limits is investigated and finally we conclude in Sec.~\ref{sec:con}.

\section{Linear sigma model coupled to quarks}
\label{lsm}
 The Lagrangian density for the LSMq reads 
\begin{align}
\mathcal{L}&=\frac{1}{2}(\partial_{\mu}\sigma)^2+\frac{1}{2}(\partial_{\mu}\bm{\bm{\pi}})^2+\frac{a^2}{2}(\sigma^2+\bm{\pi}^2) \nn &-\frac{\lambda}{4}(\sigma^2+\bm{\pi}^2)^2+i\bar{\psi}\gamma^{\mu}\partial_{\mu}\psi-g\bar{\psi}(\sigma+i\gamma_5\bm{\tau}\cdot\bm{\pi})\psi\ . \label{unb_lag}
\end{align}
The first four terms on the rhs of Eq.~\eqref{unb_lag} constitute the LSM part and the last two terms are the quark part of the Lagrangian density. Here $\bm{\pi}=(\pi^1,\pi^2,\pi^3)$. The charged and neutral pion fields are defined as
\be
\pi^{\pm} = \frac{1}{\sqrt{2}}\left(\pi^{1}\pm i \pi^2\right), \qquad \pi^0 = \pi^{3}, 
\ee
respectively, $\sigma$ is the sigma meson of the LSM, and  $\psi$ is the light quark doublet with
\be
\psi = \begin{pmatrix}
	u \\ d
\end{pmatrix},
\ee
$\bm{\tau} = (\tau^1,\tau^2,\tau^3)$, where $\tau^i$ $(i=1,2,3)$ is the $i^{th}$ Pauli spin matrix.
Also, $a^2$ is the mass parameter of the theory but, unlike in the usual convention, we take $a^2<0$ in the symmetry-unbroken state. Finally, $\lambda$ is the coupling within $\sigma$-$\sigma$, $\pi$-$\pi$, and  $\sigma$-$\pi$, $g$ is the coupling between the degrees of freedom (d.o.f) of the LSM and those of the quarks.

When $a^2>0$, the  $O(4)$ symmetry of the Lagrangian is spontaneously broken and the $\sigma$ field gets a nonzero vacuum expectation value (VEV). After the symmetry breaking, $\sigma$ becomes
\begin{align}
\sigma \rightarrow \sigma + v,
\end{align}
where $v$ is the VEV developed by the $\sigma$ field.
The Lagrangian density is written after the shift as follows:
\begin{align}
\mathcal{L} &= \bar{\psi}(i\gamma^{\mu}\partial_{\mu}-M_{\sF})\psi+\frac{1}{2}(\partial_{\mu}\sigma)^2+\frac{1}{2}(\partial_{\mu}\mathbb{\bm{\pi}})^2 \nn
&-\frac{1}{2}M^2_{\sigma}\sigma^2-\frac{1}{2}M^2_{\pi}\bm{\pi}^2 \nn
&-g\bar{\psi}(\sigma+i\gamma_5\bm{\tau}\cdot\bm{\pi})\psi -V(\sigma,\pi)-V_{\rm tree}(v), \label{eq:lag_ssb}
\end{align}
with 
\begin{align}
V(\sigma,\pi) &= \lambda v \sigma(\sigma^2+\bm{\pi}^2)+\frac{\lambda}{4} (\sigma^2+\bm{\pi}^2)^2, \label{eq:pot_V} \\
V_{\text{tree}}(v)&=-\frac{1}{2}a^2v^2+\frac{1}{4}\lambda v^4 .\label{eq:pot_tree} 
\end{align}
The masses of the quarks, three pions, and sigma are given by 
\begin{align}
M_{\sF} &= gv, \nn 
M^2_{\pi} &= \lambda v^2 - a^2, \nn
M^2_{\sigma} &= 3 \lambda v^2 - a^2. \label{eq:tree_masses}
\end{align}
Note that the minimum of the tree-level potential, obtained by solving $\,\,\displaystyle \frac{dV_{\text{tree}}(v)}{dv}\bigg\vert_{v=v_0}=0$, is given by
\begin{align}
v_{0} = \sqrt{\frac{a^2}{\lambda}}.
\end{align}
Therefore, the masses after symmetry breaking, evaluated at $v_0$, are given by
\begin{align}
M_{\sF}(v_{0}) &= gv_0, \nn
M^2_{\pi} &= 0, \nn
M^{2}_{\sigma} &= 2a^2.
\end{align}
To incorporate a nonvanishing pion mass into the model, we add a small explicit symmetry-breaking term to the Lagrangian, 
\begin{align}
\mathcal{L}\rightarrow \mathcal{L}^{\prime} = \mathcal{L}+\mathcal{L}_{ESB} = \mathcal{L} + \frac{1}{2}m_{\pi}^2v (\sigma + v) ,
\end{align}
with $m_{\pi} = 0.14 \,\,\text{GeV}$. As a result, the tree potential $V_{\text{tree}}$ becomes $V_{\text{tree}}^{\prime} = -\frac{1}{2}(a^2+m^2_{\pi})v^2+\frac{1}{4}\lambda v^4$ and the minimum is shifted to 
\begin{align}
v_{0} \rightarrow v^{\prime}_{0} = \left(\frac{a^2+m^2_{\pi}}{\lambda}\right)^{1/2}.
\end{align}
The masses, evaluated at this new minimum $v_0^{\prime}$, are given by
\begin{align}
M_{\sF}(v^{\prime}_0) &= g \left(\frac{a^2+m^2_{\pi}}{\lambda}\right)^{1/2}, \label{eq:Mf_mod} \\
M^2_{\pi}(v^{\prime}_0) &= m^{2}_{\pi}, \nn
M^2_{\sigma}(v^{\prime}_0) &= 2a^2+3m^2_{\pi}. \label{eq:mass_modified}
\end{align} 
The value of $a$ is given by solving Eq.~\eqref{eq:mass_modified} as
\begin{equation}
a=\sqrt{\frac{M^2_{\sigma}(v^{\prime}_0)-3M^2_{\pi}(v^{\prime}_0)}{2}}\simeq \sqrt{\frac{m^2_{\sigma}-3m^2_{\pi}}{2}}
\end{equation}
\\ \\
We consider a time-independent homogeneous background magnetic field in the $z$ direction $\bm{\mathcal{B}} = B \hat{\bm{z}}$, which can be obtained from the symmetric four-potential $\mathcal{A}^{\mu}=\displaystyle \frac{B}{2}(0,-y,x,0)$. As a result, for the charged d.o.f (quarks and charged pions), the four-derivative $\partial_{\mu}$ is replaced by covariant four-derivative $D_{\mu}=\partial_{\mu}+i\mathcal{Q}\mathcal{A}^{\mu}$. Here, $\mathcal{Q}=q_{\sF}$ for quarks of flavor $f$ and $\mathcal{Q}=e$ for $\pi^{\pm}$, respectively. 
\section{One-loop pion self-energy} \label{sec:pi0}
\begin{figure}[!h]
	\centering
	\includegraphics[scale=0.5]{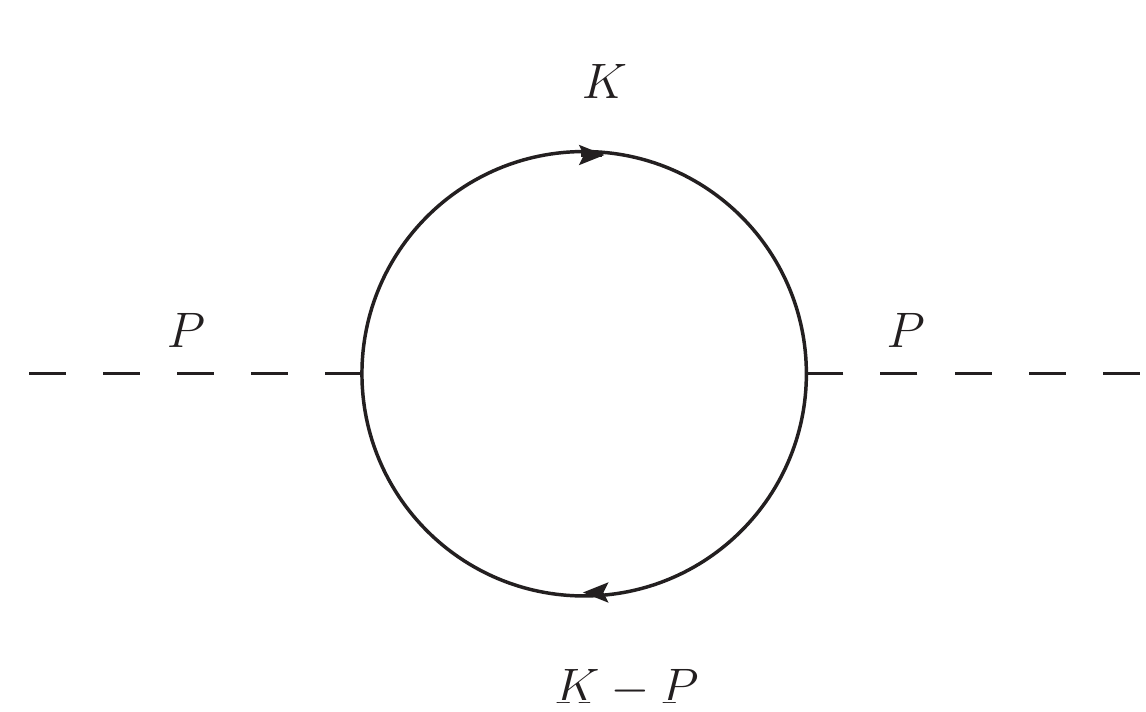}
	\caption{Feynman diagram for the one-loop quark-antiquark contribution to the $\pi^{0}$ self-energy}
	\label{fig:pi_quark}
\end{figure}

The following four terms contribute to the neutral pion self-energy:
\be
\Pi(B, P)=\Pi_{f \overline{f}}(B, P)+\Pi_{\pi^{ \pm}}(B)+\Pi_{\pi^{0}}+\Pi_{\sigma}. \label{eq:sigma_con}
\ee 
The one-loop diagram for the quark-antiquark contribution $\Pi_{f \overline{f}}(B, P)$ is depicted in Fig.~\ref{fig:pi_quark}, whereas that for the charged pion contribution $\Pi_{\pi^{ \pm}}(B)$ is depicted in Fig.~\ref{fig:pi_pi}. Note that for the last two terms $[\Pi_{\pi^{0},\sigma}]$, there are no magnetic corrections as the particles in the loop are charge neutral. In the next section, we  compute the first two contributions to the self-energy in Eq.~\eqref{eq:sigma_con}.
\subsection{Pion to quark-antiquark loop}
\label{sec:Piff}
In the presence of a magnetic field, the expression for the pion self-energy reads
\begin{align}
&\Pi_{\sF{\bar\sF}}(B,P)\nn
&=i\sum_{\sF}g^2\int\frac{d^4K}{(2\pi)^4}\text{Tr}[\gamma_5 iS_{\sF}^{\sB}(K) \gamma_5 iS_{\sF}^{\sB}(K-P)], \label{eq-ff}
\end{align}
where $S_{\sF}^{\sB}(K)$ is the charged quark propagator in momentum space in Schwinger's proper-time representation. It is given by
\begin{align}
&iS_{\sF}^{\sB}(K) \nn
&= \int_{0}^{\infty} ds\,\exp\left[is\left\{K_{\shp}^{2}+K^2_{\sperp}\frac{\tan(|q_{\sF}B|s)}{|q_{\sF}B|s}-M_{\sF}^2+i\epsilon\right\}\right]\nn &\hspace{0.5cm}\times\Big[\left(\slashed{K}_{\shp}+M_{\sF}\right)\big\{1+\textsf{sgn}(q_{\sF}B)\tan(|q_{\sF}B|s)\gamma^1\gamma^2\big\}\nn
&\hspace{4cm} + \slashed{K}_{\sperp}\sec^2(|q_{\sF}B|s)\Big], \label{eq:full_sf}
\end{align}
where $\textsf{sgn}$ is the sign function.

We follow the following notations and conventions:
\begin{itemize}
	\item Four-vectors are denoted by capital letters.
	\item The presence of a magnetic field in the $z$ direction breaks the rotational symmetry of the system. Therefore, we decompose any four-vector into its parallel and perpendicular components. Hence, for the momentum four-vector $K^{\mu}=(k^0,k^1,k^2,k^3)$, we have $K_{\sperp}^{\mu}=(0,k^1,k^2,0)$ and $K_{\shp}^{\mu}=(k^0,0,0,k^3)$. 
	\item Likewise, for any two four-vectors $A^{\mu}=(a^0,a^1,a^2,a^3)$ and $B^{\mu}=(b^0,b^1,b^2,b^3)$, we define the parallel and perpendicular dot product as $(a\cdot b)_{\shp} = A\cdot B_{\shp}=a^0b^0-a^3b^3$ and $(a\cdot b)_{\sperp}=-A\cdot B_{\sperp}=a^1b^1+a^2b^2$, respectively.
	\item Also, for any four-vector $A^{\mu}$, we use the standard Feynman slash notation $\slashed{A}$ to indicate $\gamma^{\mu}A_{\mu}$. 
\end{itemize}

After calculating the trace over Dirac matrices, Eq.~\eqref{eq-ff} becomes
\begin{align}
&\Pi_{\sF\bar{\sF}}(B,P)=4i\sum_{\sF}g^2\int\frac{d^4K}{(2\pi)^4}\int\limits_{0}^{\infty} ds\,dt\nn
&\ \times\,e^{is\left(K_{\shp}^{2}+K^2_{\sperp}\frac{\tan(|q_{\sF}B|s)}{|q_{\sF}B|s}-M_{\sF}^2\right)} e^{it\left(Q_{\shp}^{2}+Q^2_{\sperp}\frac{\tan(|q_{\sF}B|t)}{|q_{\sF}B|t}-M_{\sF}^2\right)}
\nn &\ \times\Big[\left(K\cdot Q_{\shp}-M^2_{\sF}\right)\big\{\tan(|q_{f}B|s)\tan(|q_{f}B|t)-1\big\} \nn
&\hspace{1.5cm} +\sec^{2}(|q_{f}B|s)\sec^{2}(|q_{f}B|t)\(K\cdot Q\)_{\sperp}\Big], \label{eq:Pi_qq}
\end{align}
where $Q=K-P$.
Now, to carry out the four-momentum integration in Eq.~\eqref{eq:Pi_qq}, we switch to Euclidean spacetime by the usual replacement $k^0\rightarrow ik^0_E$ and also with the substitution $( s\rightarrow -is,\ \ t\rightarrow -it)$ as in Ref.~\cite{Alexandre:2000jc}. The subscript $E$ stands for momentum components in Euclidean spacetime. Now, we get 
\begin{align}
&\Pi_{\sF\bar{\sF}}(B,P)=4\sum_{\sF}g^2\int\frac{d^4K_E}{(2\pi)^4}\int\limits_{0}^{\infty} ds\,dt\,\nn
&\times \exp\left[-s\left\{(K_E^{\shp})^2-(K_E^{\sperp})^2\frac{\tanh(|q_{\sF}B|s)}{|q_{\sF}B|s}+M_{\sF}^2\right\}\right] \nn
&\times \exp\left[-t\left\{(Q_E^{\shp})^2-(Q_E^{\sperp})^2\frac{\tanh(|q_{\sF}B|t)}{|q_{\sF}B|t}+M_{\sF}^2\right\}\right]\nn
&\times \Big[\left\{M^2_{\sF}+(K\cdot Q)_E^{\shp}\right\}\big\{1+\tanh(|q_{f}B|s)\tanh(|q_{f}B|t)\big\}\nn
&\hspace{1cm}+\sech^{2}(|q_{f}B|s)\sech^{2}(|q_{f}B|t)(K\cdot Q)_E^{\sperp}\Big],
\label{eq:euc_Pi_qq}
\end{align}
where we defined 
\be
d^4K_E &\equiv& dk_E^0\,dk_E^1\,dk_E^2\,dk_E^3,\nn
d^2K^{\shp}_E &\equiv& dk_E^0\,dk_E^3,\nn
d^2K^{\sperp}_E &\equiv& dk_E^1\,dk_E^2,\nn
(K_E^{\shp})^2 &\equiv& (k_E^0)^2+(k_E^3)^2,\nn
(K_E^{\sperp})^2 &\equiv& -(k_E^{\sperp})^2 \equiv -\left\{(k_E^1)^2+(k_E^2)^2\right\}, \nn
(K\cdot Q)_E^{\shp} &\equiv& k_E^0q_E^0+k_E^3q_E^3, \nn
(K\cdot Q)_E^{\sperp} &\equiv& -\left\{k_E^1q_E^1+k_E^2q_E^2\right\}.
\ee
 Note that three-momentum in Minkowski and Euclidean space are the same. We also used the identities $\tan(-i x)=-i\tanh(x)$ and $\sec(-i x)=\sech(x)$. \\
The momentum integration in Eq.~\eqref{eq:euc_Pi_qq} can be performed analytically: 
\begin{align}
&\Pi_{\sF\bar{\sF}}(B,P)= 4\sum_{f}g^2\int_{0}^{\infty}ds\,dt\nn
&\hspace{0.5cm}\times\Bigg[\bigg(1+\tanh(|q_{f}B|s)\tanh(|q_{f}B|t)\bigg)\mathcal{I}_{\shp}^{(1)}\mathcal{I}_{\sperp}^{(0)}\nn
&\hspace{1.5cm}+\, \sech^{2}(|q_{f}B|s)\sech^{2}(|q_{f}B|t)\mathcal{I}_{\shp}^{(0)}\mathcal{I}_{\sperp}^{(1)}\Bigg],
\end{align} 
where 

\be
&&\hspace{-.5cm}\mathcal{I}_{\shp}^{(0)}(s,t,P_E^{\shp},M_{\sF})\nn
&&\hspace{.5cm}\equiv\int\frac{d^2K^{\shp}_E}{(2\pi)^2} e^{-[s(K_E^{\shp})^2+t(Q_E^{\shp})^2+M^2_{\sF}(s+t)]} \nn
&&\hspace{.5cm}=\frac{e^{-\big\{M^2_{\sF}(s+t)+(P_E^{\shp})^2\frac{st}{s+t}\big\}}}{4\pi(s+t)},  \label{eq:I0paral}\\
&&\hspace{-.5cm}\mathcal{I}_{\shp}^{(1)}(s,t,P_E^{\shp},M_{\sF})\equiv\int\frac{d^2K^{\shp}_E}{(2\pi)^2}\left\{M^2_{\sF}+(K\cdot Q)_E^{\shp}\right\}\nn
&&\hspace{1.5cm}\times\ e^{-[s(K_E^{\shp})^2+t(Q_E^{\shp})^2+M^2_{\sF}(s+t)]}\nn
&=&\hspace{.5cm}\frac{e^{-\big\{M^2_{\sF}(s+t)+(P_E^{\shp})^2\frac{st}{s+t}\big\}}}{4\pi (s+t)^2}\nn
&&\hspace{1.5cm}\times\left[1+M^2_{\sF}(s+t)-(P_E^{\shp})^2\frac{st}{s+t}\right], \label{eq:I1paral}
\ee
\be
&&\hspace{-.5cm}\mathcal{I}_{\sperp}^{(0)}(s,t,P_E^{\sperp},|q_{\sF}B|)\\
&&\equiv \int\frac{d^2K^{\sperp}_E}{(2\pi)^2}e^{\frac{(K_E^{\sperp})^2\tanh(|q_{\sF}B|s)+(Q_E^{\sperp})^2\tanh(|q_{\sF}B|t)}{|q_{\sF}B|}} \nn
&&=\frac{|q_{\sF}B|\exp\left[\frac{(P_E^{\sperp})^2}{|q_{\sF}B|}\frac{\sinh(|q_{\sF}B|s)\sinh(|q_{\sF}B|t)}{\sinh(|q_{\sF}B|(s+t))}\right]}{4\pi[\tanh(|q_{\sF}B|s)+\tanh(|q_{\sF}B|t)]}, \label{eq:I0perp}
\ee
\be
&&\hspace{-.5cm}\mathcal{I}_{\sperp}^{(1)}(s,t,p_E^{\sperp},|q_{\sF}B|)\equiv \int\frac{d^2K^{\sperp}_E}{(2\pi)^2}\ (K\cdot Q)_E^{\sperp}  \nn
&&\hspace{1.5cm}\times\ e^{\frac{(K_E^{\sperp})^2\tanh(|q_{\sF}B|s)+(Q_E^{\sperp})^2\tanh(|q_{\sF}B|t)}{|q_{\sF}B|}}\nn &=&\frac{|q_{\sF}B|^2\exp\left[\frac{(P_E^{\sperp})^2}{|q_{\sF}B|}\frac{\sinh(|q_{\sF}B|s)\sinh(|q_{\sF}B|t)}{\sinh(|q_{\sF}B|(s+t))}\right]}{4\pi[\tanh(|q_{\sF}B|s)+\tanh(|q_{\sF}B|t)]^2}\nn
&\times&\Bigg\{1+\frac{(P_E^{\sperp})^2}{|q_{\sF}B|}\frac{\sinh(|q_{\sF}B|s)\sinh(|q_{\sF}B|t)}{\sinh(|q_{\sF}B|(s+t))}\Bigg\}. \label{eq:I1perp}
\ee
Thus, we are left with only two proper-time integrations in the expression of $\Pi_{\sF\bar{\sF}}$ :
\begin{align}
&\Pi_{\sF\bar{\sF}}(B,P) = \sum_{\sF}\frac{g^2}{4\pi^2}\int\limits_{0}^{\infty}ds\,dt\,\frac{|q_{\sF}B|}{(s+t)}\nn
&\times e^{-\left\{M^2_{\sF}(s+t)+(P_E^{\shp})^2\frac{st}{s+t}-\frac{(P_E^{\sperp})^2}{|q_{\sF}B|}\frac{\sinh(|q_{\sF}B|s)\sinh(|q_{\sF}B|t)}{\sinh[|q_{\sF}B|(s+t)]}\right\}} \nn
&\hspace{-0cm}\times\Bigg[\frac{1+M^2_{\sF}(s+t)-(P_E^{\shp})^2\frac{st}{s+t}}{\(s+t\)\tanh\(|q_fB|(s+t)\)}-\frac{|q_fB|}{\sinh^2\(|q_fB|(s+t)\)} \nn
&\times\Bigg(1+\frac{(P^{\sperp}_E)^2}{|q_fB|}\frac{\sinh(|q_{\sF}B|s)\sinh(|q_{\sF}B|t)}{\sinh[|q_{\sF}B|(s+t)]}\Bigg)\Bigg]. \label{eq:full_mom_Piff}
\end{align}
Since we are interested in calculating the pion mass modification given by Eq.~\eqref{eq:ex_disp}, we take the limit $\bm{p}\rightarrow \bm{0}$ of Eq.~\eqref{eq:full_mom_Piff} to get
\begin{align}
&\hspace{-0.5cm}\Pi_{\sF\bar{\sF}}(B,p_E^0)\nn
&=\sum_{\sF}\frac{g^2}{4\pi^2}\int\limits_{0}^{\infty}ds\,dt\,\frac{|q_{\sF}B|e^{-(s+t)M^2_{\sF}-(p^0_E)^2\frac{s t}{s+t}}}{(s+t)}\nn
&\times\Bigg[\frac{1+M^2_{\sF}(s+t)-\frac{s t}{s+t}(p_E^0)^2}{(s+t)\tanh\big(|q_{\sF}B|(s+t)\big)} \nn
&\hspace{2cm}-\frac{|q_{\sF}B|}{\sinh^2\big(|q_{\sF}B|(s+t)\big)}\Bigg]. \label{eq:Piffp0E}
\end{align}
In the vanishing magnetic field limit, $\Pi_{\sF\bar{\sF}}(B,p_E^0)$ becomes
\begin{align}
\Pi_{\sF\bar{\sF}}(B\rightarrow 0,p_E^0) &= \sum_{\sF}\frac{g^2}{4\pi^2}\int\limits_{0}^{\infty}ds\,dt\,\frac{e^{-(s+t)M^2_{\sF}-(p^0_E)^2\frac{s t}{s+t}}}{(s+t)}\nn
&\times\frac{M^2_{\sF}(s+t)-\frac{s t}{s+t}(p_E^0)^2}{(s+t)^2}. 
 \label{eq:PiffeB0}
\end{align}
After rotating back to Minkowski spacetime, Eq.~\eqref{eq:PiffeB0} can also be expressed in terms of a four-momentum integration as
\begin{align}
&\Pi_{\sF\bar{\sF}}(B\rightarrow 0,p^0) =-4i\sum_{\sF}g^2\int\frac{d^4K}{(2\pi)^4}\nn
&\hspace{1cm}\times \frac{k_0\(k_0-p_0\) - M_{\sF}^2}{\(K^2-M_{\sF}^2\)\(\(k_0-p_0\)^2-k^2-M_{\sF}^2\)}.
 \label{eq:PiffeB02}
\end{align}
Equation~\eqref{eq:PiffeB02} is UV divergent and can be computed using the Feynman parametrization. The integrand can be regularized using the $\overline{\rm MS}$ scheme and the result will contribute to the renormalized vacuum pion mass. As we are interested in the magnetic correction to the pion mass, we define the Vacuum part subtracted self-energy as
\begin{align}
&\Pi_{\sF\bar{\sF}}(B,p_E^0)\nn 
&\hspace{0.2cm}=\sum_{\sF}\frac{g^2}{4\pi^2}\int\limits_{0}^{\infty}ds\,dt\,\frac{|q_{\sF}B|e^{-(s+t)M^2_{\sF}-(p^0_E)^2\frac{s t}{s+t}}}{(s+t)}\nn
&\hspace{0.2cm}\times\Bigg[\frac{1+M^2_{\sF}(s+t)-\frac{s t}{s+t}(p_E^0)^2}{(s+t)\tanh\big(|q_{\sF}B|(s+t)\big)} \nn
&\hspace{0.cm}-\frac{|q_{\sF}B|}{\sinh^2\big(|q_{\sF}B|(s+t)\big)}-\frac{M^2_{\sF}(s+t)-\frac{s t}{s+t}(p_E^0)^2}{|q_{\sF}B|(s+t)^2}\Bigg]. 
\label{eq:Piffp0}
\end{align}
\subsection{Charged pion loop}
\label{sec:Pipm}
\begin{figure}[!h]
	\centering
	\includegraphics[scale=0.5]{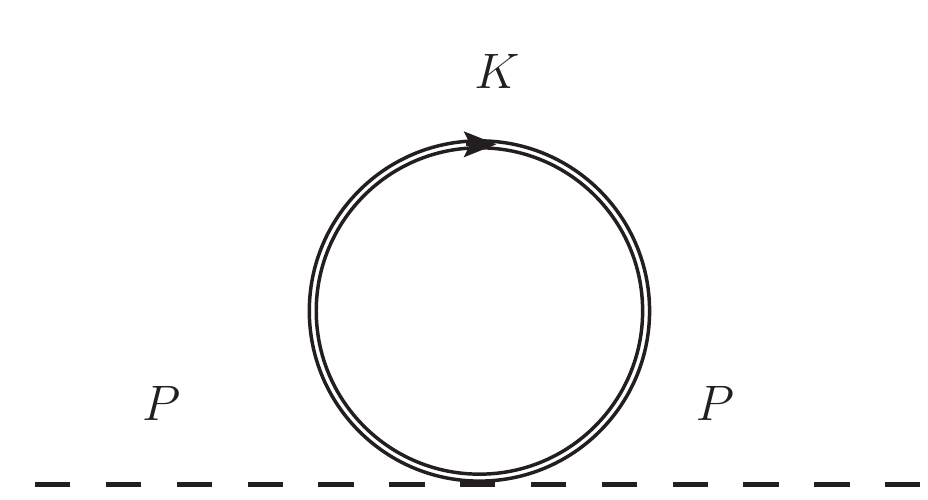}
	\caption{Feynman diagram for the one-loop charged pion contribution to the $\pi^{0}$ self-energy}
	\label{fig:pi_pi}
\end{figure}
The tadpole diagram, shown in Fig.~\ref{fig:pi_pi}, reads
\begin{align}
\Pi_{\pi_{\pm}}(B)=\frac{\lambda}{4}\int\frac{d^4K}{(2\pi)^4}iD_{B}(K). \label{eq:pi_pi}
\end{align}
where $D^{(B)}(k)$ is the charged pion propagator in the presence of a magnetic field, given by
\begin{align}
iD^{(B)}(K) =\int\limits_{0}^{\infty}\frac{ds}{\cos(|eB| s)}e^{is\left(K^2_{\shp}+K^2_{\sperp}\frac{\tan(|eB|s)}{|eB|s}-m_{\pi}^2+i\epsilon\right) }. \label{eq:scalar_prop}
\end{align}
We go to Euclidean spacetime and get
\begin{align}
\Pi_{\pi_{\pm}}(B)=\frac{\lambda}{4}\int\limits_{0}^{\infty}\frac{ds}{\cosh(|eB|s)}\mathcal{J}_{\shp}^{(0)}\mathcal{J}_{\sperp}^{(0)}, \label{eq:pi_pi_intermediate}
\end{align}
where $\mathcal{J}_{\shp}^{(0)}$ and $\mathcal{J}_{\sperp}^{(0)}$ are given by
\begin{align}
\mathcal{J}_{\shp}^{(0)}(s,m^2_{\pi})& = \int\frac{d^2K_E^{\shp}}{(2\pi)^2} e^{-\left((K_E^{\shp})^2+m^2_{\pi}\right)s }, \label{eq:J0_paral}\\
\mathcal{J}_{\sperp}^{(0)}(s,|eB|)& = \int\frac{d^2K_E^{\sperp}}{(2\pi)^2} e^{(K_E^{\sperp})^2\frac{\tanh(|eB|s)}{|eB|}}. \label{eq:J0_perp}
\end{align}
 The momentum integrations in Eqs.~\eqref{eq:J0_paral} and~\eqref{eq:J0_perp} can be performed analytically to obtain
 \begin{align}
 \mathcal{J}_{\shp}^{(0)}(s,m^2_{\pi})& = \frac{e^{-s\,m^2_{\pi}}}{4\pi s}, \label{eq:J0_paral_intg}\\
 \mathcal{J}_{\sperp}^{(0)}(s,|eB|)& =\frac{|eB|}{4\pi\tanh(|eB|s)}. \label{eq:J0_perp_intg}
 \end{align}
 Plugging Eqs.~\eqref{eq:J0_paral_intg}~\eqref{eq:J0_perp_intg} into  Eq.~\eqref{eq:pi_pi_intermediate}, we get
\begin{align}
\Pi_{\pi_{\pm}}(B) = \frac{\lambda}{4}\frac{|eB|}{16\pi^2}\int\limits_{0}^{\infty} \frac{ds}{s} \frac{e^{-sm^2_{\pi}}}{\sinh(|eB|s)}. \label{eq:pi_pi_complete}
\end{align}
For small $eB$, we can series expand the integrand in Eq.~\eqref{eq:pi_pi_complete} around $eB=0$ and carry out the integration over the proper time $s$. The terms with odd powers of $eB$ do not appear since $\Pi_{\pi_{\pm}}(B)$ is even in $|eB|$. In the vanishing magnetic field limit, $\Pi_{\pi_{\pm}}(B)$ becomes
\begin{align}
\Pi_{\pi_{\pm}}(B\rightarrow 0) = \frac{\lambda}{4}\frac{1}{16\pi^2}\int\limits_{0}^{\infty}ds \frac{e^{-sm^2_{\pi}}}{s^2}.
\label{eq:pi_pieB0}
\end{align}
Equation~\eqref{eq:pi_pieB0} can also be expressed in terms of an integration over four-momentum 
from Eq.~\eqref{eq:pi_pi} taking $eB\rightarrow0$ as
\begin{align}
\Pi_{\pi_{\pm}}(B\rightarrow 0)=\frac{\lambda}{4}\int\frac{d^4K}{(2\pi)^4}\frac{i}{K^2-m_\pi^2+i\epsilon}. \label{eq:pi_pieB02}
\end{align}
Equation~\eqref{eq:pi_pieB02} is divergent and can be renormalized using the $\overline{\rm MS}$ scheme and contribute to the renormalized vacuum pion mass. As in Sec.~\ref{sec:Piff}, we define the vacuum part subtracted self-energy as
\begin{align}
&\Pi_{\pi_{\pm}}(B)\nn
&=\frac{\lambda}{4}\frac{1}{16\pi^2}\int\limits_{0}^{\infty}ds\frac{e^{-sm^2_{\pi}}}{s}\left[\frac{|eB|}{\sinh(|eB|s)}-\frac{1}{s}\right].\label{eq:pi_pi_p0E}
\end{align} 
\vspace{0.0cm}
\section{Pion Mass} \label{sec:Pi_mass}
In order to compute the modified pion mass $M_{\pi}(B)$ in the presence of a magnetic field, we need to solve the equation
\begin{equation}
p^2_0-|\bm{p}|^2-m^2_{\pi}-\text{Re}[\Pi(B,P)]=0 \label{eq:ex_disp}
\end{equation}
in the limit $\bm{p}\rightarrow 0$ and $p_0=M_{\pi}(B)$. The self-energy of $\pi^0$ has four contributions as mentioned in Eq.~\eqref{eq:sigma_con}.
Now, $\Pi_{f \overline{f}}(B, P)$ and $\Pi_{\pi_{\pm}}(B)$ were calculated in Secs.~\ref{sec:Piff} and~\ref{sec:Pipm}, respectively. But $\Pi_{\pi^0}(P)$ and $\Pi_{\sigma}(P)$ do not receive any magnetic field corrections since the particles in the loop are neutral.
So, the total self-energy $\Pi(B,p_0,\bm{p}=\bm{0})$ is obtained from Eqs.~\eqref{eq:Piffp0E} and~\eqref{eq:pi_pi_p0E} as 
\begin{align}
&\Pi(B,p_0)=\sum_f \frac{g^2}{4\pi^2}\int\limits_{0}^{\infty}ds\,dt\,\frac{|q_{\sF}B|}{(s+t)}e^{-(s+t)M^2_{\sF}-\frac{s t}{s+t}(p^0_E)^2}\nn
&\hspace{0.2cm}\times\Bigg[\frac{1+M^2_{\sF}(s+t)-\frac{s t}{s+t}(p^0_E)^2}{(s+t)\tanh\(|q_{\sF}B|(s+t)\)}\nn
&\hspace{.20cm}-\frac{|q_{\sF}B|}{\sinh^2\(|q_{\sF}B|(s+t)\)} -\frac{M_{\sF}^2(s+t)-\frac{s t}{s+t}(p^0_E)^2}{|q_{\sF}B|\(s+t\)^2}\Bigg] \nn
&\hspace{0.2cm} + \frac{\lambda}{4}\frac{1}{16\pi^2}\int\limits_{0}^{\infty}ds\frac{e^{-sm^2_{\pi}}}{s}\left[\frac{|eB|}{\sinh(|eB|s)}-\frac{1}{s}\right]. \label{eq:Pi_ex_st}
\end{align}
We can make a change of variable in Eq.~\eqref{eq:Pi_ex_st} from $(s,t)$ to $(u, v)$ (as done in Refs.~\cite{Alexandre:2000jc,Tsai:1974ap})
\begin{align}
s=\frac{1}{2}u(1-v),\qquad t=\frac{1}{2}u(1+v),
\end{align} 
which leads to
\begin{align}
&\Pi(B,p_0) = \sum_{f} \frac{\lambda}{4\pi^2}\int\limits_{0}^{\infty}du\int\limits_{-1}^{1}dv\,\frac{M_f^2|q_{\sF}B|}{2\(a^2+m_\pi^2\)}\nn
&\quad\times\exp\left[-u\left(M^2_{\sF}+\frac{1}{4}(1-v^2)(p_E^0)^2\right)\right]\nn
&\quad\times\left[\frac{1+uM^2_{\sF}-\frac{1}{4}u(1-v^2)(p_E^0)^2}{u\tanh(|q_{\sF}B|u)} \right. \nn
&\quad\left.-\frac{|q_{\sF}B|}{\sinh^{2}(|q_{\sF}B|u)}-\frac{M^2_{\sF}-\frac{1}{4}(1-v^2)(p_E^0)^2}{|q_{\sF}B|u}\right]\nn
&\quad+\frac{\lambda}{4}\frac{1}{16\pi^2}\int\limits_{0}^{\infty}du\frac{e^{-u\,m^2_{\pi}}}{u}\left[\frac{|eB|}{\sinh(|eB|u)}-\frac{1}{u}\right]. \label{eq:Pi_ex_uv}
\end{align}
Using Eq.~\eqref{eq:Pi_ex_uv}, the solution of $p_0$ from Eq.~\eqref{eq:ex_disp} gives the transcendental equation for the magnetic-field-dependent neutral pion mass as
\begin{widetext}
\be
M_{\pi}^2(B) &=&  (v_0^{\prime})^2\lambda-a^2 -  \sum_{f} \frac{g^2}{4\pi^2}\int\limits_{0}^{\infty}du\int\limits_{-1}^{1}dv\,\frac{|q_{\sF}B|\, }{2}e^{-u\left[M_{\sF}^2-\frac{1}{4}(1-v^2)M_{\pi}^2(B)\right]}\left[\left\lbrace\frac{1+uM^2_{\sF}+\frac{1}{4}u(1-v^2)M^2_{\pi}(B)}{u\tanh(|q_{\sF}B|u)} \right.\right. \nn
&&\left.\left.-\frac{|q_{\sF}B|}{\sinh^{2}(|q_{\sF}B|u)}\right\rbrace-\frac{M_{\sF}^2+\frac{1}{4}(1-v^2)M^2_{\pi}(B)}{u}\right]-\frac{\lambda}{4}\frac{1}{16\pi^2}\int\limits_{0}^{\infty}du\frac{e^{-u\,m^2_{\pi}}}{u}\left[\frac{|eB|}{\sinh(|eB|u)}-\frac{1}{u}\right].
\label{MpiB}
\ee
Equation~\eqref{MpiB} can be solved numerically to find the magnetic-field dependent pion mass, shown by the blue line Fig.~\ref{fig:eff_comp}.
\begin{figure}[tbh]
	\centering
	\includegraphics[scale=0.43]{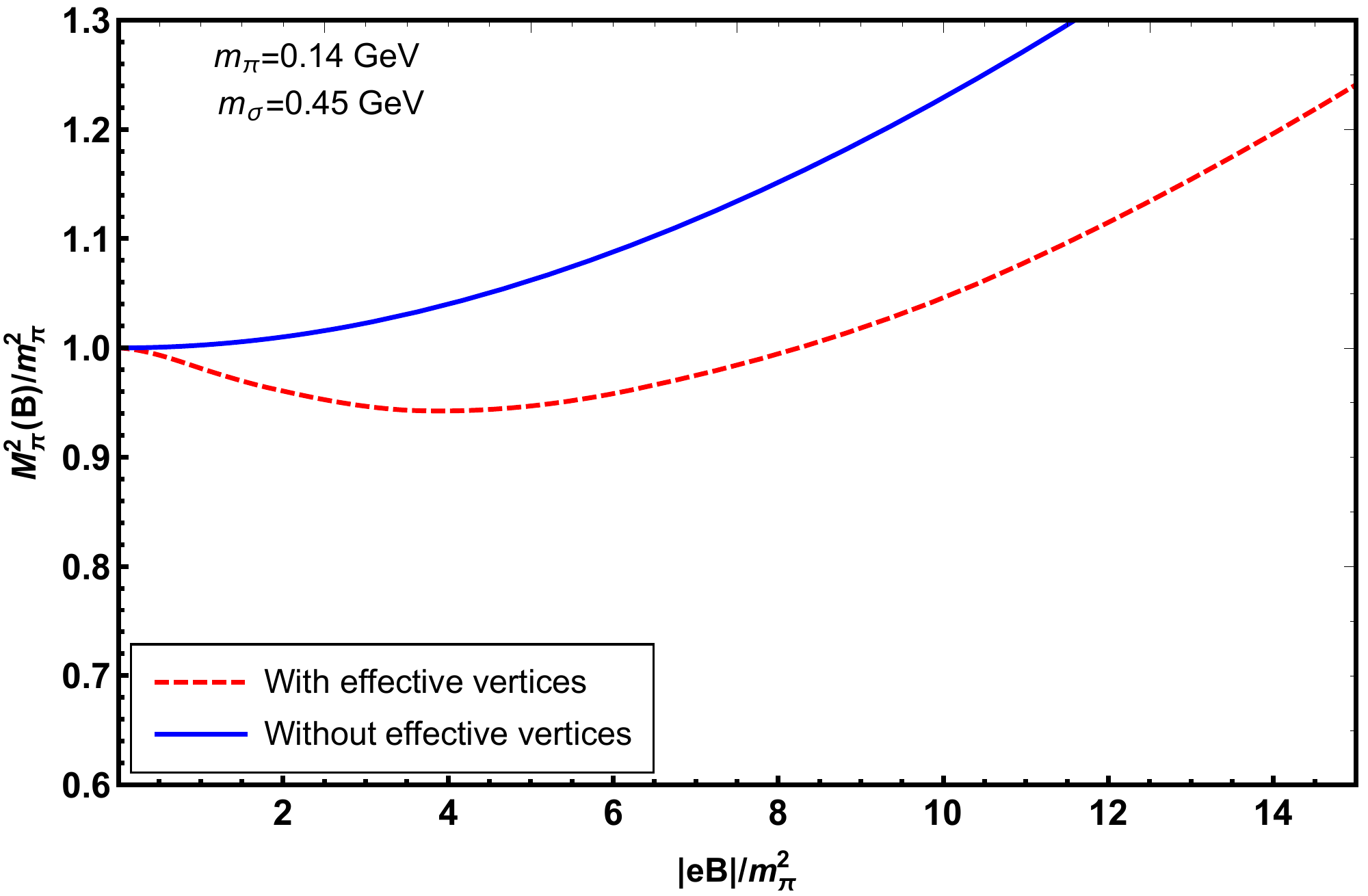}
	\caption{The magnetic-field-dependent neutral pion mass from Eqs.~\eqref{MpiB} and~\eqref{MpiB_eff}.}
	\label{fig:eff_comp}
\end{figure}
As is clear from the Fig.~\ref{fig:eff_comp}, the magnetic-field-dependent neutral pion mass increases with the field strength. But the calculation is not complete yet: we also need to incorporate the one-loop magnetic-field correction to the fermion coupling $g$, the boson self-coupling $\lambda$ and $v_0'$. 

So, the effective fermion mass becomes
\be
M_{\sF,\rm{eff}} &= g_{\rm eff}\, v_0^{B},
\label{eff_quantities} 
\ee
where $v_0^{B}$ represents the magnetic-field-dependent minimum of the potential after symmetry breaking and $g_{\rm eff}$ is the one-loop effective fermion vertex in the presence of the magnetic field. 

Using Eq.~\eqref{eff_quantities} and replacing $g_{\rm eff}$ with the other effective quantities, Eq.~\eqref{MpiB} becomes
\be
M_{\pi}^2(B) &=&  (v_0^{B})^2\lambda_{\text{eff}}-a^2 -  \sum_{f} \frac{1}{4\pi^2}\frac{1}{(v_0^{B})^2}\int\limits_{0}^{\infty}du\int\limits_{-1}^{1}dv\,\frac{|q_{\sF}B|\, M_{\sF,\rm{eff}}^2}{2}e^{-u\left[M_{\sF}^2-\frac{1}{4}(1-v^2)M_{\pi}^2(B)\right]}\nn
&\times&\left[\left\lbrace\frac{1+uM^2_{\sF}+\frac{1}{4}u(1-v^2)M^2_{\pi}(B)}{u\tanh(|q_{\sF}B|u)}  -\frac{|q_{\sF}B|}{\sinh^{2}(|q_{\sF}B|u)}\right\rbrace-\frac{M_{\sF}^2+\frac{1}{4}(1-v^2)M^2_{\pi}(B)}{u}\right]\nn
&-&\frac{\lambda_{\rm eff}}{4}\frac{1}{16\pi^2}\int\limits_{0}^{\infty}du\frac{e^{-u\,m^2_{\pi}}}{u}\left[\frac{|eB|}{\sinh(|eB|u)}-\frac{1}{u}\right].
\label{MpiB_eff}
\ee
The effective magnetic-field-dependent quantities, namely,  $v_0^{B}$, $\lambda_{\rm eff}$ and $M_{\sF,\rm{eff}}$ are calculated in Appendices~\ref{sec:v0B},~\ref{sec:vertex} and ~\ref{sec:q_mass}, respectively. The solution for $M_{\pi}(B)$ from Eq.~\eqref{MpiB_eff} for fixed values of $m_{\sigma}$ and $m_{\pi}$ are plotted in the left and right panels of Fig.~\ref{fig:pion_mass_exact_full}, respectively.
\begin{figure}[tbh]
	\centering
\subfigure{
	\includegraphics[scale=0.42]{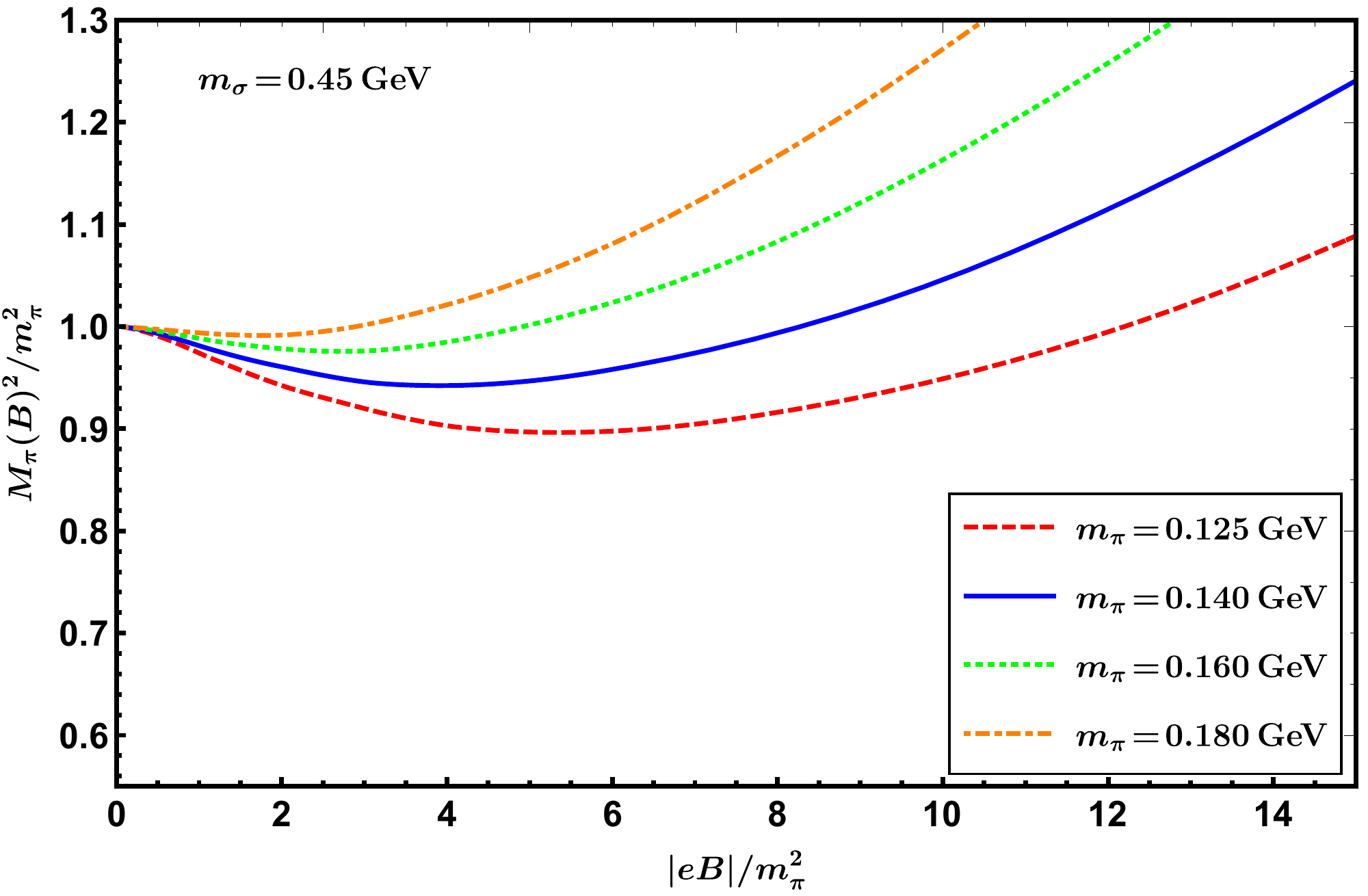}}
\subfigure{
	\includegraphics[scale=0.42]{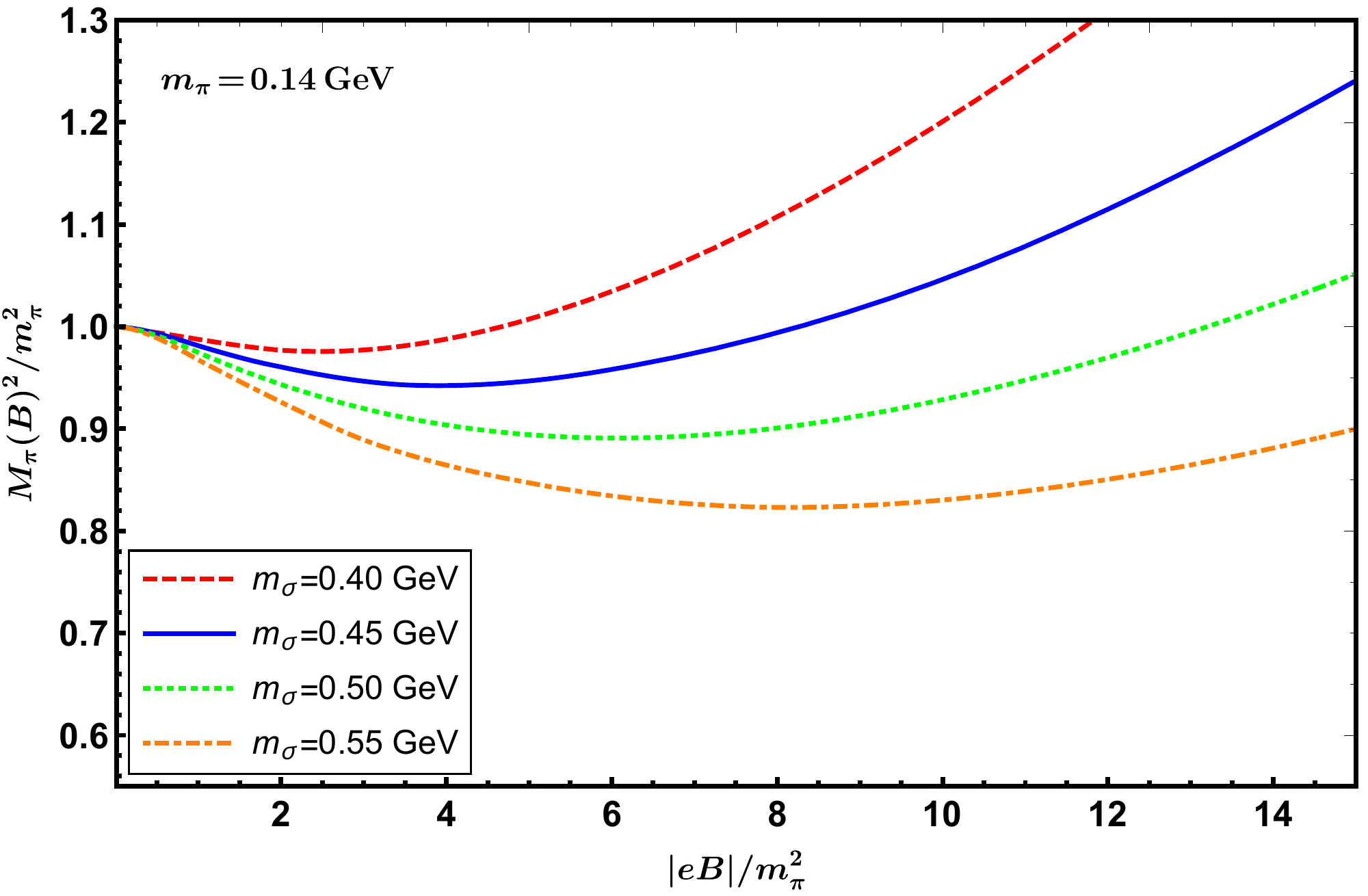}}
	\caption{The magnetic field dependence of the neutral pion mass for a fixed $m_{\sigma}=0.45\,\,\text{GeV}$ with $m_{\pi}=0.125,\, 0.140,\, 0.160,\, 0.180 \,\text{GeV} $ (left panel) and for a fixed $m_{\pi}=0.14\,\, \text{GeV}$ with $m_{\pi}=0.40,\, 0.45,\, 0.50,\, 0.55\,\, \text{GeV} $ (right panel). The coupling constants $\lambda=0.86$ and $g=1.11$ are taken from Ref.~\cite{Ayala:2015hba}.}
	\label{fig:pion_mass_exact_full}
\end{figure}
\end{widetext}
Figure~\ref{fig:pion_mass_exact_full} shows the nonmonotonic behavior of neutral pion mass with magnetic field. It~\emph{decreases} with increasing magnetic field for a weak magnetic field in agreement with the Ref.~\cite{Ayala:2018zat}. But for large values of magnetic field, the neutral pion mass starts to increase. This could be due to the fact that the linear sigma model coupled to quarks is best at mimicking the low-energy region of QCD, but in the high-energy region (such as for large magnetic fields) this model is not reliable.
\section{Asymptotic solution}\label{sec:asymp}
It is possible to find analytic expressions in asymptotic limits such as the weak-field limit as done in Ref.~\cite{Ayala:2018zat}. Expanding the integrand in Eq.~\eqref{eq:Piffp0E} in powers of $q_{\sF}B$ and performing the $u$ and $v$ integrations analytically order by order in $q_{\sF}B$, we get the vacuum -subtracted contribution from quark-loop in weak field as  
\begin{align}
&\Pi^{w}_{\sF\bar{\sF}}(B,p_E^0) =\sum_{\sF}\frac{g^2}{4\pi^2} \nn
& \times\left[\frac{4}{3}\frac{(q_{\sF}B)^2}{(p^0_E)^2+4 \(M_{\sF,\rm{eff}}^w\)^2}\Bigg(1\right. +\frac{1}{p^0_E}\frac{(p^0_E)^2+8  \(M_{\sF,\rm{eff}}^w\)^2}{(p^0_E)^2+4  \(M_{\sF,\rm{eff}}^w\)^2}\nn
&\hspace{0.3cm}\left.\times\sinh^{-1}\left[\frac{p_E^0}{2 M_{\sF,\rm{eff}}^w}\right]\Bigg)-\frac{2(q_{\sF}B)^4}{45\left[(p^0_E)^2+4M_{\sF}^2\right]^3} \right.\nn
&\left.\times\left\{\frac{24}{p_E^0}\frac{(p^E_0)^2+24 M_{\sF}^2}{(p^E_0)^2+4 M_{\sF}^2}\sinh^{-1}\left[\frac{p^0_E}{2 M_{\sF}}\right]+304\right.\right. \nn
&\hspace{1.5cm}\left.\left.+\ 66\frac{(p_E^0)^2}{M_{\sF}^2}+5\frac{(p_E^0)^4}{M_{\sF}^4}\right\}+\cdots\right],\label{eq:ff_weak_all}
\end{align}
where $M_{\sF,\rm{eff}}^w$ represents weak-field expression of the quark mass up to $\mathcal{O}(eB)^2$. The coefficient of $\mathcal{O}(eB)^2$ in Eq.~\eqref{eq:ff_weak_all} is different than that obtained in Ref.~\cite{Ayala:2018zat} as the authors of Ref.~\cite{Ayala:2018zat} used an approximation in which they considered $p_0^E=0$. If we make the same approximation, we get the vacuum-subtracted contribution from the charged pion loop in weak field as
\begin{align}
&\Pi^{w}_{\sF\bar{\sF}}(B,0)\nn
& =\sum_{\sF}\frac{g^2}{4\pi^2}\(\frac{(q_{\sF}B)^2}{\(M_{\sF,\rm{eff}}^w\)^2}-\frac{14}{45}\frac{(q_{\sF}B)^4}{M_{\sF}^6}+\cdots\).
\end{align}
This result matches that in Ref.~\cite{Ayala:2018zat} at $\mathcal{O}[(q_{\sF}B)^2]$.
In a similar manner, in the weak magnetic field limit the charged pion tadpole diagram is obtained by expanding Eq.~\eqref{eq:pi_pi_p0E} in the small-$eB$ limit and doing the proper-time integration, which gives
\begin{align}
&\Pi_{\pi_{\pm}}^w(B) \nn
&=\frac{\lambda}{64 \pi^2}\left[ -\frac{(eB)^2}{6m^2_{\pi}}+\frac{7(eB)^4}{180m^6_{\pi}}-\frac{31(eB)^6}{360m^{10}_{\pi}}+ ...\right]. \label{eq:pi_pi_weak_all}
\end{align}
Note that the $\mathcal{O}(eB)^2$ term matches that in Ref.~\cite{Ayala:2018zat}.
In weak-field limit, the transcendental Equation~\eqref{eq:Pi_ex_uv} can be solved for $M^2_{\pi}(B)$ and we get
\begin{align}
\left.M^2_{\pi}(B)\right|_w &= \lambda_{\rm eff}^w {v^{\prime}_0}^2-a^2 \nn
&+\sum_f \frac{\lambda_{\rm eff}M_{\sF}^2 (q_{\sF}B)^2}{\pi^2\(m_\pi^2 + a^2\)} \left\{\frac{1}{3 \left(4M_{\sF}^2+m_\pi^2\right)}\right.\nn
&\left.+\frac{\left(8M_{\sF}^2+m_\pi^2\right) }{3m_\pi \left(4 M_{\sF}^2+m_\pi^2\right)^{3/2}}\sinh^{-1}\left(\frac{m_\pi}{2M_{\sF}}\right)\right\} \nn
&-\frac{\lambda_{\rm eff}^w}{64 \pi^2} \frac{(eB)^2}{6m^2_{\pi}}+\mathcal{O}\(eB\)^4 ,
\label{Mpi_B_w1}
\end{align}
where we have replaced $M_{\sF,\rm{eff}}^w$ with $M_{\sF}$ as the magnetic field correction in $M_{\sF,\rm{eff}}^w$ in the weak-field limit is $\mathcal{O}\(eB\)^2$. We have also approximated the magnetic-field-dependent minimum of the potential as $v_0^B\approx v_0'$ in Eq.~\eqref{Mpi_B_w1}. Using $\lambda_{\rm eff}^w$ from Eq.~\eqref{lambda_eff_w}, Eq.~\eqref{Mpi_B_w1} becomes
\be
\left. M^2_{\pi}(B)\right|_w &=&m_\pi^2+\sum_f \frac{\lambda M_{\sF}^2 (q_{\sF}B)^2}{\pi^2\(m_\pi^2 + a^2\)} \left\{\frac{1}{3 \left(4M_{\sF}^2+m_\pi^2\right)}\right.\nn
&&\left.\hspace{-1.5cm} +\ \frac{\left(8M_{\sF}^2+m_\pi^2\right) }{3m_\pi \left(4 M^2+m_\pi^2\right)^{3/2}}\sinh^{-1}\left(\frac{m_\pi}{2M_{\sF}}\right)\right\} - \frac{\lambda}{8 \pi^2} \frac{(eB)^2}{m^2_{\pi}} \nn
&&\hspace{-1.5cm} \times\left\{\frac{1}{48} +\frac{a^2+m_{\pi}^2}{5m^2_{\pi}}\(1+\frac{4\csch^{-1}2}{\sqrt{5}}\) \right\}+ \mathcal{O}\(eB\)^4. \qquad
\label{Mpi_B_w}
\ee
\begin{figure}[!h]
	\centering
	\includegraphics[scale=0.42]{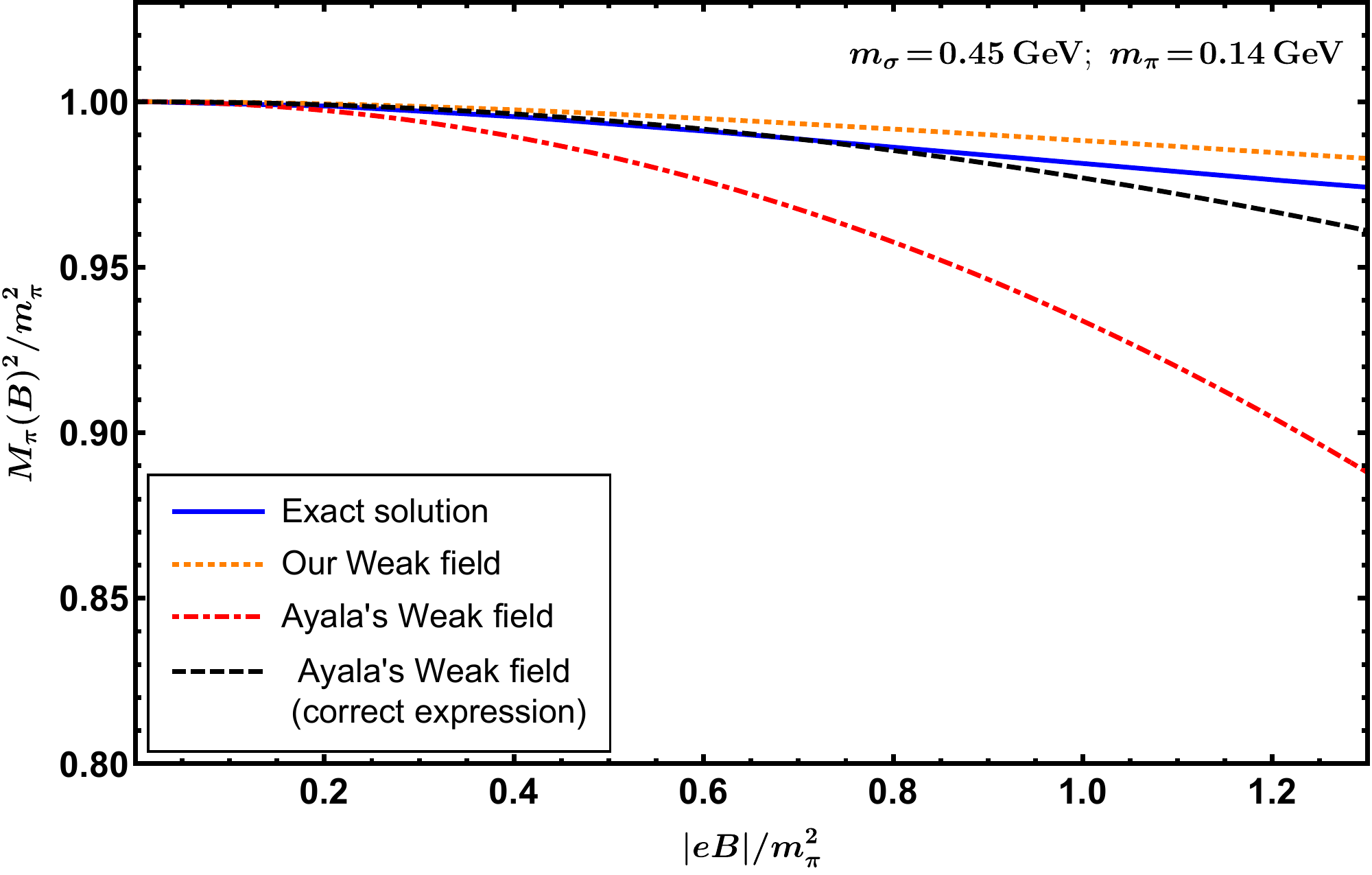}
	\caption{Comparison of the weak-field result [Eq.~\ref{Mpi_B_w}] with (a) the exact solution, (b) the weak-field result from Ref.~\cite{Ayala:2018zat}, and (c) the corrected weak-field result of Ref.~\cite{Ayala:2018zat} as given in Eq.~\eqref{eq:ayala_corrected} .}
	\label{fig:weak_comp}
\end{figure}

\section{Conclusion and outlook} \label{sec:con}
In conclusion, we have studied modifications of neutral pion mass in the presence of an external background magnetic field under the framework of the LSMq. Our calculation is valid for weak to moderate external background magnetic fields. We also obtained an asymptotic solution of the dispersion equation of the neutral pion mass in the weak magnetic field limit. The calculation was carried out by taking into account the self-coupling of pions as well as the effective one-loop quark in the presence of a magnetic field. It matches that of Ayala \textit{et al.}~\cite{Ayala:2018zat} in the low$-|eB|$ regime. But, when the strength of the magnetic field is increased, we get a nonmonotonic behavior, unlike in Ref.~\cite{Ayala:2018zat}. Up to a moderate value of the magnetic field, our result also qualitatively agrees with LQCD studies as~\cite{Bali:2017ian} in that the pion mass decreases with increasing magnetic field for low $|eB|$. Additionally, it is clear from Fig.~\ref{fig:weak_comp} that the pion mass in a weak magnetic field as given in Eq.~\eqref{Mpi_B_w} is a good approximation for $eB\lesssim m_\pi^2$. Moreover, the corrected weak-field result from Ref.~\cite{Ayala:2018zat} [as given in Eq.~\eqref{eq:ayala_corrected}] is more or less the same as our weak-field pion mass [Eq.~\ref{Mpi_B_w}], which was obtained without approximating $p_0=0$ in the dispersion relation in Eq.~\eqref{eq:ex_disp} up to $|eB|\lesssim 0.85\, m^2_{\pi}$. For $|eB|\gtrsim 0.85\, m^2_{\pi}$, the corrected weak-field expression [as given in Eq.~\eqref{eq:ayala_corrected}] starts to deviates and our weak-field expression is more closer to the exact value of the magnetic-field-dependent pion mass.

Looking to the future, the present calculation can be extended to astrophysical objects, where the baryon density and magnetic field are very high. We can also extend the present calculation at finite temperature, which could be interesting for heavy-ion physics. Finally, one should keep in mind that the LSM is an effective model and it has a scalar degree of freedom (called sigma meson) which has not been confirmed yet but is still under consideration after the identification of $f_0(500)$ with $J^{PC}=0^{++}$. The nonlinear sigma model(NLSM) was proposed as an extension of the LSM long ago~\cite{Koch:1997ei,Koch:1995vp}, where the $\sigma$ field is removed by sending its mass to infinity and redefining the pion field $\bm{\Phi}(x)$ by $U(x)=\exp(i\bm{\tau}\cdot\bm{\Phi}(x)/f_{\pi})$; the study of light mesons in the presence of a magnetic field in the NLSM might be interesting. Nevertheless, using the LSMq model, we can qualitatively capture the essential features obtained by much more involved and rigorous studies.    

\section{Acknowledgments}
The authors would like to acknowledge M. G. Mustafa, Arghya Mukherjee and Pradip K. Roy for useful discussions and careful reading of the article.  A. D. was funded by the Department of Atomic Energy (DAE), India via the Saha Institute of Nuclear Physics and partially by the School of Physical Sciences, National Institute of Science Education and Research (NISER). N. H. was funded by DAE/NISER.

\appendix
\section{Effective potential and magnetic-field-dependent VEV}
\label{sec:v0B}
The one-loop effective potential is
\be
V_{\rm{eff}} = V_{\rm{cl}} + V_{\rm{b}} + V_{\rm{f}} + V_{c}, 
\label{v_eff}
\ee
where the tree-level potential after symmetry breaking is
\be
V_{\rm{cl}}=-\frac{\left(a^{2}+m_{\pi}^{2}\right)}{2} v^{2}+\frac{\lambda}{4} v^{4}.
\ee
The one-loop effective potential for the boson fields is
\be
V_{b}&=& \sum_{b=\sigma,\pi^0,\pi^{\pm}}\frac{1}{2} \int dm_{b}^{2} \int\frac{d^{4} K_E}{(2 \pi)^{4}} \int\limits_{0}^{\infty} \frac{d s}{\cosh \left(|q_{b} B| s\right)} \nn
  & \times& e^{-s\left((k_E^{0})^{2}+k_{3}^{2}+k_{\perp}^{2} \frac{\tanh \left(|q_{b} B| s\right)}{q_{b} B s}+m_{b}^{2}\right)}
\ee 
The one-loop effective potential for the fermion fields is
\be
 V_{f}&=&-\sum_{r=\pm 1}\sum_{f=u,d}  \int d m_{f}^{2} \int \frac{d^{4} K_E}{(2 \pi)^{4}} \int_{0}^{\infty} \frac{d s}{\cosh \left(|q_{f}B| s\right)} \nn
  & \times& e^{-s\left((k_E^{0})^{2}+k_{3}^{2}+k_{\perp}^{2} \frac{\tanh \left(|q_{f}B|s\right)}{|q_{f}B| s}+m_{f}^{2}+r |q_{f}B|\right)}
\ee  
After calculating the integrals, the renormalized potentials become
\begin{widetext}
\be
V_{f}&=&-2\left\{\frac{(g v)^{4}}{16 \pi^{2}}\left[\ln \left(\frac{(g v)^{2}}{a^{2}}\right)+\frac{1}{2}\right]\right\} \nn
 &+&\frac{1}{8 \pi ^2}\sum_f\left[ \frac{3}{4} \left((gv)^4-a^4\right)+4 q_fB^2 \left\{\psi _{-2}\left(\frac{(gv)^2}{2 |q_fB|}\right)-\psi^{(-2)}\left(\frac{a^2}{2 |q_fB|}\right)\right\}-|q_fB| (1+\log (2 \pi )) \big\{(gv)^2-a^2\big\}\right.\nn
 &+&\left.\frac{(gv)^2 \big\{(gv)^2-2 |q_fB|\big\}}{2}  \log \left(\frac{2 |q_fB|}{(gv)^2}\right)-\frac{a^2}{2}  \left(a^2-2 |q_fB|\right) \ln\left(\frac{2 |q_fB|}{a^2}\right)\right] 
\ee
and
\be
V_{b}^{1}&=&3\left\{\frac{\left(\lambda v^{2}-a^{2}\right)^{2}}{64 \pi^{2}}\left[\ln \left(\frac{\lambda v^{2}-a^{2}}{a^{2}}\right)+\frac{1}{2}\right]\right\}
+\left\{\frac{\left(3 \lambda v^{2}-a^{2}\right)^{2}}{64 \pi^{2}}\left[\ln \left(\frac{3 \lambda v^{2}-a^{2}}{a^{2}}\right)+\frac{1}{2}\right]\right\}\nn
&+&\frac{2}{128 \pi ^2} \left[3 \big\{(\lambda v^2-a^2)^2-a^4\big\}-4 |eB| \log (2 \pi ) \left(\lambda v^2-2a^2\right)-2 a^4 \ln \left(\frac{2 |eB|}{a^2}\right)+2 (\lambda v^2-a^2)^2 \ln \left(\frac{2|eB|}{\lambda v^2-a^2}\right)\right.\nn
&+&16 eB^2 \left\{\psi^{(-2)}\left(\frac{\lambda v^2-a^2+|eB|}{2 |eB|}\right)-\psi^{(-2)}\left(\frac{a^2+|eB|}{2 |eB|}\right)\right\}\Bigg],
\ee
where $\psi^m(z)$ represents the derivative of the logarithm of the gamma function as  $\psi^m(z)=\frac{d^{m+1}}{dz^{m+1}}\ln\Gamma(z).$ 
\begin{figure}[tbh]
	\subfigure{
\includegraphics[scale=0.64]{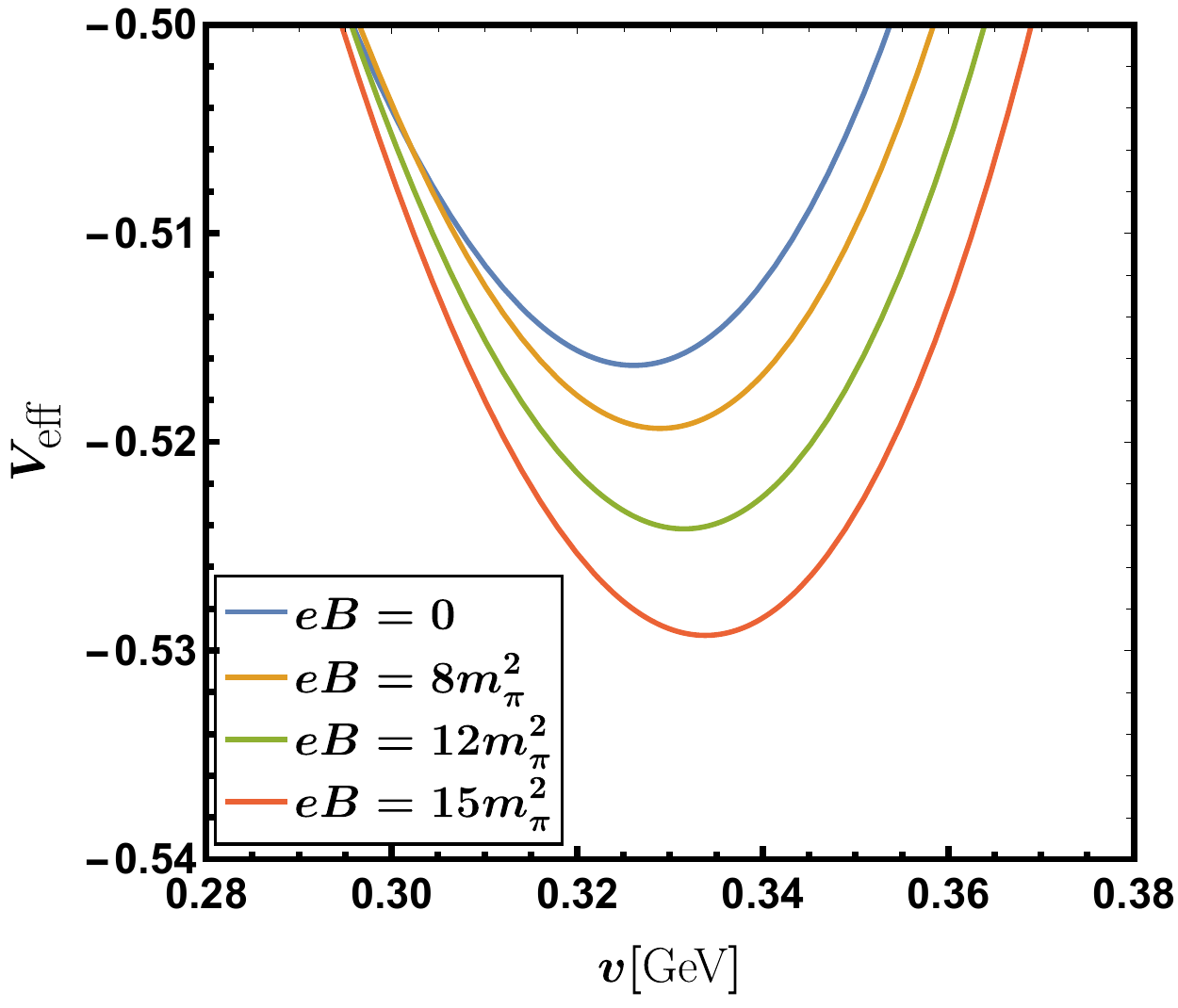}}
	\subfigure{
	\includegraphics[scale=0.60]{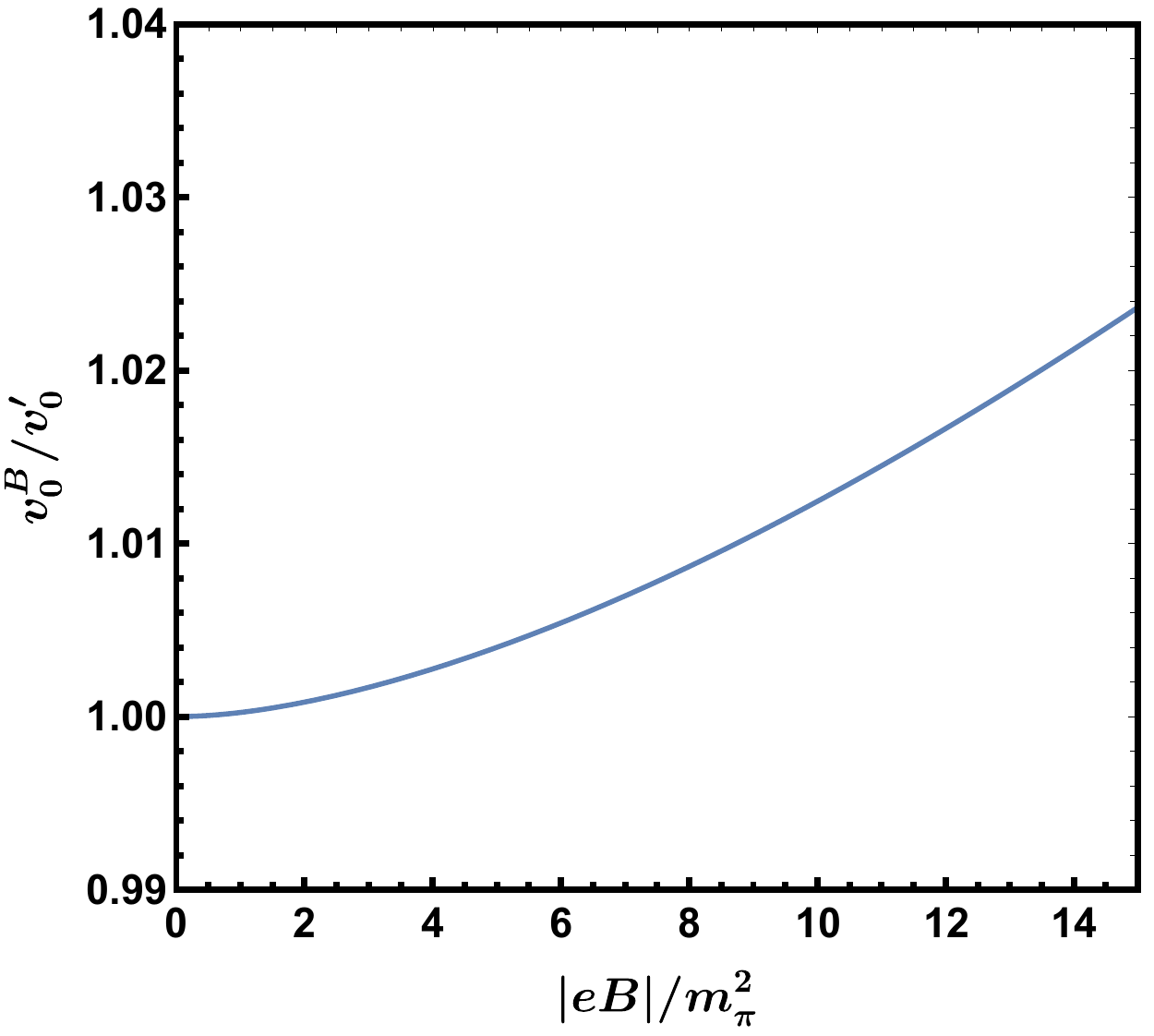}}
\caption{Left: Magnetic-field-dependent effective potential. Right: Magnetic-field-dependent minimum of the effective potential}
\label{fig:v0B}
\end{figure}
\end{widetext}
The total effective potential can be obtained from Eq.~\eqref{v_eff} by adding the individual contributions. In the left panel of Fig.~\ref{fig:v0B} we plot the effective potential as a function of the VEV $v$. It is clear from the figure that the position of the minimum varies with $v$. In the right panel of Fig.~\ref{fig:v0B} we plot the scaled VEV as a function of the magnetic field.

\section{Effective boson self-coupling vertex}
\label{sec:vertex}
\begin{figure}[!h]
	\centering
	\includegraphics[scale=0.5]{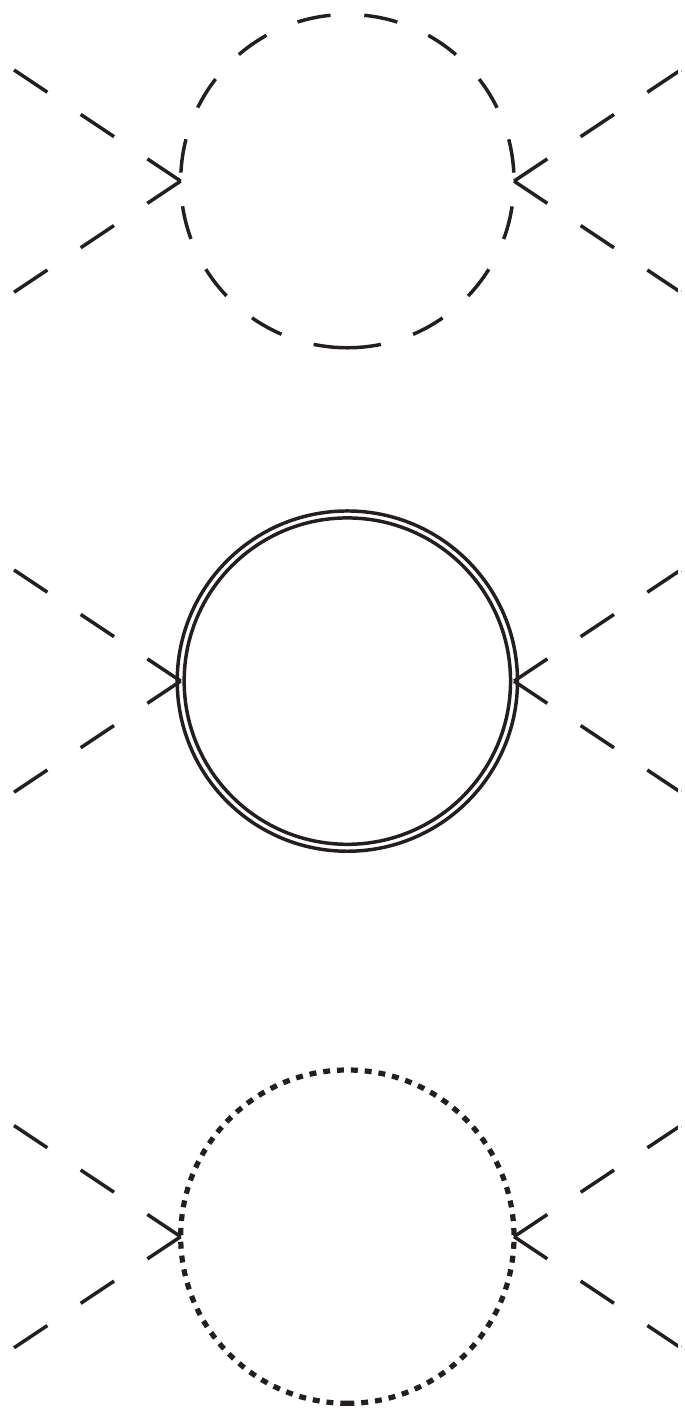}
	\caption{Feynman diagram for the magnetic field correction to the self-coupling $\lambda$. The dashed line denotes $\pi^{0}$, the double line denotes $\pi^{\pm}$, and the dotted line denotes the $\sigma$ meson.}
	\label{fig:pi_vertex}
\end{figure}

In this section, we calculate the boson self-coupling up to one-loop order, as shown in the diagram in Fig.~\ref{fig:pi_vertex}.

The effective vertex up to one-loop order is written as
\begin{align}
\lambda_{\text{eff}}=\lambda+\Delta\lambda,
\end{align}
where $\Delta\lambda$ is given by~\cite{Ayala:2014gwa}
\begin{align}
\Delta\lambda = \frac{24\lambda^2}{16}\left[9I(B,P,m_{\sigma})+I(B,P,m_{\pi})\right. \nn
\left.\hspace{4 cm}+\ 4J(B,P,m_{\pi})\right]\Bigg|_{\bm{p}\rightarrow \bm{0}}. \label{eq:delta_lambda}
\end{align}
Here 
\begin{align}
I(B,P,m_{i}) &= -i\int\frac{d^4K}{(2\pi)^4}D_{i}(K)D_{i}(P-K), \label{neutral_I} \\ \nn
J(B,P,m_j) &= -i\int\frac{d^4K}{(2\pi)^4}D^{B}_{j}(K)D^{B}_{j}(P-K), \label{charged_J}
\end{align}
with $i=\pi^0, \sigma$ and $j=\pi^{+},\pi^{-}$. Equations~\eqref{neutral_I} and \eqref{charged_J} are one-loop contributions from neutral and charged boson fields, respectively. Here $D_i(K)$ is the neutral boson propagator,
\begin{align}
D_{i}(K) = \frac{1}{K^2-M^2_i}
\end{align}
and the charged boson propagator is given by Eq.~\eqref{eq:scalar_prop}.

Now, $I(B,P,m_{i})$ does not contribute to the magnetic field corrections to the boson self-coupling, but $J(B,P,m_j)$ does. Converting Eqs.~\eqref{eq:scalar_prop} and~\eqref{charged_J} to Euclidean space, we have
\begin{align}
&J(B,p_E^0,m_j)\nn
&= \int\frac{d^4K_E}{(2\pi)^4}D_j^B(K_E)D_j^B(K_E-P_E) \nn
&=\int\limits_{0}^{\infty}ds\,dt\,\frac{\mathcal{I}_{\sperp}^{(1)}\mathcal{I}_{\shp}^{(0)} }{\cosh(|eB|s)\cosh(|eB|t)},
\label{JB_st}
\end{align}
where $\mathcal{I}_{\shp}^{(0)}$ and $\mathcal{I}_{\sperp}^{(0)}$ are given by Eqs.~\eqref{eq:I0paral} and~\eqref{eq:I0perp}, respectively, with $M_{\sF}$ replaced by $m_{\pi}$. 

Now, using Eqs.~\eqref{eq:I0paral} and \eqref{eq:I0perp}, Eq.~\eqref{JB_st} becomes
\begin{align}
&\hspace{-0.1cm}J(B,P_E,m_\pi) =\frac{|eB|}{16\pi^2} \int\limits_{0}^{\infty}\frac{ds\,dt}{(s+t) \sinh\(|eB|(s+t)\)}\nn
&\hspace{-0.09cm}\times e^{-\left\{m^2_{\pi}(s+t)+(P^{\shp}_{E})^2\frac{st}{s+t} - \frac{(P^{\sperp}_{E})^2}{|eB|}\frac{\tanh(|eB|s)\tanh(|eB|t)}{\tanh(|eB|s)+\tanh(|eB|t)}\right\}}. \label{eq:Jp_E}
\end{align} 
In the limit that the three-momentum $\bm{p}\rightarrow \bm{0}$ and after changing variables from $(s,t)$ to $(u,v)$, Eq.~\eqref{eq:Jp_E} is further simplified to
\begin{align}
&\hspace{-0.5cm}J(B,p^0_E,m_\pi)\nn
&=\frac{1}{16\pi^2}\int\limits_{-1}^{1}dv\int\limits_{0}^{\infty}du\,\frac{|eB|e^{-u\,\(m^2_{\pi}+\frac{1}{4}(1-v^2)(p_E^0)^2\)}}{2\sinh\big(|eB|u\big)}. \label{eq:Jp0E_uv}
\end{align}
In the vanishing magnetic field limit, Eq.~\eqref{eq:Jp0E_uv} becomes
\be
&&\hspace{-1cm}J(B\rightarrow 0,p^0_E,m_\pi)\nn
&=&\frac{1}{16\pi^2}\int\limits_{-1}^{1}dv\int\limits_{0}^{\infty}du\,\frac{e^{-u\,\(m^2_{\pi}+\frac{1}{4}(1-v^2)(p_E^0)^2\)}}{2u}.
\label{J_eB0}
\ee 
Equation~\eqref{J_eB0} can also be expressed in terms of the momentum integration from Eq.~\eqref{JB_st} as
\be
&&\hspace{-0.6cm}J(B\rightarrow 0,p^0_E,m_\pi)\nn
&=& \int\frac{d^4K_E}{(2\pi)^4}\frac{1}{K_E^2+m_\pi^2}\frac{1}{\(p_E^0-k_E^0\)^2 + \mathbf{k}^2+m_\pi^2},\qquad
\ee
which is UV divergent and can be regularized using the $\overline{\rm MS}$ scheme. The regularized vanishing magnetic field contribution to the vertex will only contribute to the zero-magnetic-field pion mass. As before, we define the vacuum-subtracted self-interaction vertex as
\be
&&\hspace{-0.8cm}J(B,p^0_E,m_\pi)\nn
&=&\frac{1}{16\pi^2}\int\limits_{-1}^{1}dv\int\limits_{0}^{\infty}du\,e^{-u\,\left\{m^2_{\pi}+\frac{1}{4}(1-v^2)(p_E^0)^2\right\}}\nn
&&\qquad\times\ \left[\frac{|eB|}{2\sinh\big(|eB|u\big)}-\frac{1}{2u}\right].
\label{J_ren}
\ee
Thus, from Eq.~\eqref{eq:delta_lambda} we see that the term $\Delta\lambda$ receives corrections from the magnetic field through $J$ only, and it is given by
\begin{align}
\Delta\lambda &= 6\lambda^2 J(B,p^0_E,m_\pi)\nn
&=\frac{6\lambda^2}{16\pi^2}\int\limits_{-1}^{1}dv\int\limits_{0}^{\infty}du\,e^{-u\,\left\{m^2_{\pi}+\frac{1}{4}(1-v^2)(p_E^0)^2\right\}}\nn
&\qquad\times\ \left[\frac{|eB|}{2\sinh\big(|eB|u\big)}-\frac{1}{2u}\right],
\end{align}
which leads to the one-loop neutral pion self-coupling vertex,
\be
\lambda_{\rm eff}&=& \lambda + 6\lambda^2 J(B,p^0_E,m_\pi)\nn
&=&\lambda + \frac{3\lambda^2}{8\pi^2}\int\limits_{-1}^{1}dv\int\limits_{0}^{\infty}du\,e^{-u\,\left\{m^2_{\pi}+\frac{1}{4}(1-v^2)(p_E^0)^2\right\}}\nn
&&\qquad\times\ \left[\frac{|eB|}{2\sinh\big(|eB|u\big)}-\frac{1}{2u}\right].
\ee
In the weak magnetic field limit, the integrations over $u$ and $v$ in Eq.~\eqref{J_ren} can be done analytically as
\be
&&J^w(B,p^0_E,m_\pi)\nn
&=&\frac{1}{16\pi^2}\left[- \frac{(eB)^2}{4m^2_{\pi}+(p^0_E)^2} \left\{ \frac{1}{3m^2_{\pi}}+\frac{4}{3}\frac{\sinh^{-1}\(\frac{p^0_E}{2m_{\pi}}\)}{p_E^0\sqrt{4m^2_{\pi}+(p^0_E)^2}}\right\}\right. \nn
&+&\left.\frac{7(eB)^4}{90M_\pi p^0_E(4m^2_{\pi}+(p^0_E)^2)^3}\left\{\frac{66p^0_E}{m_{\pi}}+13\(\frac{p^0_E}{m_{\pi}}\)^3\right.\right.\nn
&+&\left.\left.\(\frac{p^0_E}{m_{\pi}}\)^5 + \frac{120m_\pi\sinh^{-1}\(\frac{p^0_E}{2m_{\pi}}\)}{\sqrt{4m^2_{\pi}+(p^0_E)^2}}\right\}\right] + \mathcal{O}[eB]^6,
\label{Jhat_w}
\ee 
and we can write the weak magnetic field effective vertex as
\be
\lambda_{\rm eff}^w&=& \lambda +6\lambda^2 J^w(B,p^0_E,m_\pi).
\label{lambda_eff_w}
\ee

In the limit $p_E^0\rightarrow 0$, Eq.~\eqref{Jhat_w} becomes
\be
&&\hspace{-1.0cm}J^w(B,p^0_E\rightarrow 0,m_\pi)\nn
& =& -\frac{1}{16\pi^2}\frac{(eB)^2}{6m^{4}_{\pi}}
\(1-\frac{7}{10}\frac{(eB)^2}{6m^{4}_{\pi}}+\cdots\) .
\label{eq:ver_intgd_expan_result}
\ee
It is worth mentioning here that the first term in Eq.~\eqref{eq:ver_intgd_expan_result} differs from the corresponding expression in Ref.~\cite{Ayala:2018zat}. We have checked the calculation in Ref.~\cite{Ayala:2018zat} and we are able to get our expression as given in Eq.~\eqref{eq:ver_intgd_expan_result}. In Ref.~\cite{Ayala:2018zat}, there is a mistake during the evaluation of the integral in Eq.~(C6) over the Feynman parameter $x$. So, in light of assumptions that were used in Ref.~\cite{Ayala:2018zat}, the correct expression for the self-coupling would be
\begin{align}
\lambda^{\text{eff}}_{\text{Ayala}}=\lambda \(1-\frac{\lambda}{16 \pi^2}\frac{(eB)^2}{m_{\pi}^4}\),
\end{align}
which leads to the following expression for the effective mass:
\begin{align}
	M^2_{\pi}(B)_{\rm Ayala}\sim m^{2}_{\pi}&\left\{1-\frac{\lambda (eB)^2}{4\pi^2 m^4_{\pi}}\left[\frac{1}{4}\(1+\frac{a^2}{m^{2}_{\pi}}\) \right.\right.\nn
	&\left.\left.-\(\frac{5/9}{1+\frac{a^2}{m^{2}_{\pi}}}-\frac{1}{96}\)\right]\right\}. \label{eq:ayala_corrected}
\end{align}
\section{Effective constituent quark mass $M_{\sF, \rm{eff}}$}
\label{sec:q_mass}
In this appendix, we evaluate magnetic field correction to the constituent quark mass up to one-loop order. The pole of the one-loop effective quark propagator in the $p_0$ plane in the limit $\mathbf{p}\rightarrow 0$ gives the effective quark mass $M_{\sF, \rm{eff}}$. Now, the effective quark propagator $S^{*}_{\sF}(P)$ is related to the quark self-energy $\Sigma_{\sF}(B,P)$ through the Dyson-Schwinger equation,
\begin{align}
{S^{*}_{\sF}}^{-1}(P)={S^{(B)}_{\sF}}^{-1}(P)-\Sigma_{\sF}(B,P), 
\end{align}\\ 
where $f=u, d$. 
\begin{widetext}
	\begin{center}
		\begin{figure}[!h]
			\includegraphics[scale=0.7]{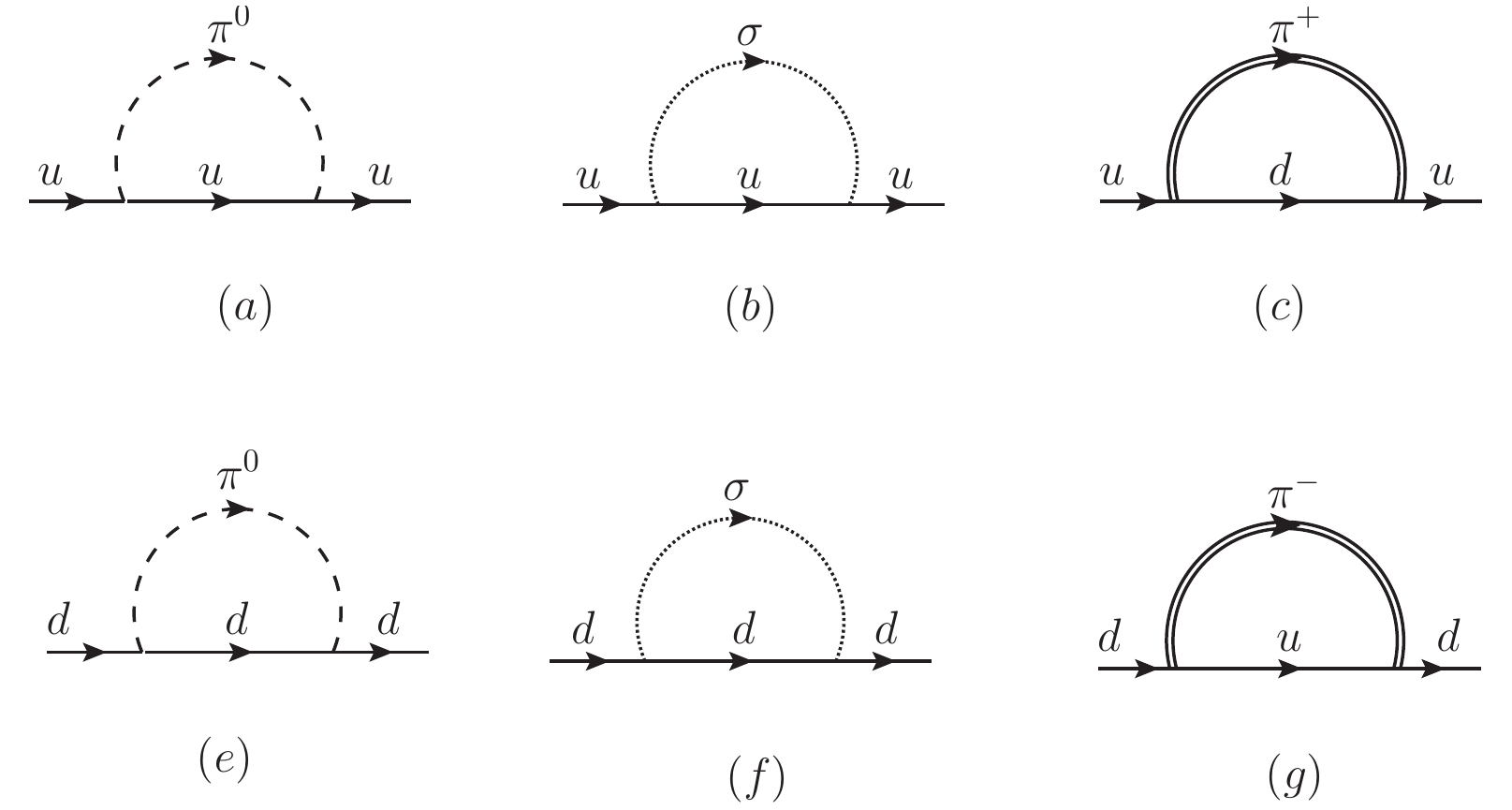}
			\caption{The Feynman diagrams that contribute to the quark self-energy.}
			\label{fig:quark_self}
		\end{figure}
	\end{center}
\end{widetext}
\subsection{Structure coefficients of ${S^{*}_{\sF}}^{-1}$}
The interaction term in the Lagrangian density, which contributes to the effective constituent quark mass, is given as 
\begin{align}
\mathcal{L}_{\sF} = -g\bar{\psi}\(\sigma+i\gamma_5\bm{\tau}.\bm{\pi}\)\psi. \label{eqn:Lf}
\end{align} 
Expanding the matrix structure in Eq.~\eqref{eqn:Lf} and simply writing it in terms of the LSM bosons ($\pi^{+}$, $\pi^{-}$, $\pi^{0}$, and $\sigma$) and constituent light quark ($u$ and $d$) fields, we get
\begin{align}
\mathcal{L}_{\sF}&= -g\(\sigma\bar{u}u+\sigma\bar{d}d+i\pi^0\bar{u}\gamma_5u-i\pi^0\bar{d}\gamma_5d\right.\nn
&\hspace{2cm}\left.+\sqrt{2}i\pi^-\bar{u}\gamma_5 d+\sqrt{2}i\pi^+\bar{d}\gamma_5 u\).
\end{align}
The contributions to the self-energy from $u$ and $d$ quarks are given in Figs.~\ref{fig:quark_self}(a)-\ref{fig:quark_self}(c) and Figs.~\ref{fig:quark_self}(e)-\ref{fig:quark_self}(g), respectively. The self-energy of the $u$ quark, $\Sigma_{u}^{(B)}$, has three contributions: $\Sigma_{[u\pi^0]}$ [Fig.~\ref{fig:quark_self}(a)], $\Sigma_{[u\sigma]}$ [Fig.~\ref{fig:quark_self}(b)], $\Sigma_{[d\pi^+]}$ [Fig.~\ref{fig:quark_self}(c)]. The self-energy of the $d$ quark, $\Sigma_d^{(B)}$, also has three contributions: $\Sigma_{[d\pi^0]}$ [Fig.~\ref{fig:quark_self}(e)], $\Sigma_{[d\sigma]}$ [Fig.~\ref{fig:quark_self}(f)], and $\Sigma_{[d\pi^-]}$ [Fig.~\ref{fig:quark_self}(g)]. This leads to 
\begin{align}
\Sigma_{u} &= \Sigma_{[u\pi^0]}+\Sigma_{[u\sigma]}+\Sigma_{[u\pi^+]}, \label{self_u} \\
\Sigma_{d} &= \Sigma_{[d\pi^0]}+\Sigma_{[d\sigma]}+\Sigma_{[d\pi^-]} \label{self_d}.
\end{align}
Each of the contributions is written separately as follows
\begin{align}
\Sigma_{[u\pi^0]}(B,P)&=-ig^2\int\dfrac{d^4K}{(2\pi)^4}\nn
&\times\left[\gamma_5S_u^{(B)}(K)\gamma_5D_{\pi^0}(P-K)\right], \label{self_[u0]}\\
\Sigma_{[u\sigma]}(B,P)&= ig^2\int\dfrac{d^4K}{(2\pi)^4}S_u^{(B)}(K)D_{\sigma}(P-K), \label{self_[us]}\\
\Sigma_{[d\pi^+]}(B,P)&=-2ig^2\int\dfrac{d^4K}{(2\pi)^4}\nn
&\times\left[\gamma_5S_d^{(B)}(K)\gamma_5D^{(B)}_{\pi^{\pm}}(P-K)\right], \label{self_[d+]}\\
\Sigma_{[d\pi^0]}(B,P)&=-ig^2\int\dfrac{d^4K}{(2\pi)^4}\nn
&\times\left[\gamma_5S_d^{(B)}(K)\gamma_5D_{\pi^0}(P-K)\right], \label{self_[d0]}\\
\Sigma_{[d\sigma]}(B,P)&= ig^2\int\dfrac{d^4K}{(2\pi)^4}\nn
&\times \left[S_d^{(B)}(K)D_{\sigma}(P-K)\right], \label{self_[ds]}\\
\Sigma_{[u\pi^+]}(B,P)&=-2ig^2\int\dfrac{d^4K}{(2\pi)^4}\nn
&\times\left[\gamma_5S_d^{(B)}(K)\gamma_5D^{(B)}_{\pi^{\pm}}(P-K)\right]. \label{self_[u-]}
\end{align} 
Now, using the anticommutator $\gamma_5\gamma^{\mu}+\gamma^{\mu}\gamma_5=0$, the term $\gamma_5S_{f}^{(B)}(K)\gamma_5$ is simplified as
\begin{align}
&\gamma_5iS_{f}^{(B)}(K)\gamma_5 =\nn
& \int\limits_{0}^{\infty} ds \exp\left[is\(K^2_{\shp} + \dfrac{\tan(|q_{\sF}B|s)}{|q_{\sF}B|s)}K^2_{\sperp}-M^2_{\sF}+i\epsilon\)\right] \nn
&\times\left[(M_{\sF}-\slashed{K}_{\shp})\{1+\textsf{sgn}(q_{\sF}B)\tan(|q_{\sF}B|s)\gamma^1\gamma^2\}\right. \nn
&\left.-\slashed{K}_{\sperp}\sec^2(|q_{\sF}B|s)\right]=iS_{f}^{(B)}(-K).
\end{align}
Substituting Eqs.~\eqref{self_[u0]}-\eqref{self_[u-]} into  Eqs.~\eqref{self_u} and~\eqref{self_d}, we get
\begin{align}
&\Sigma_u(B,P) = ig^2\int \frac{d^4K}{(2\pi)^4} \left[S^{(B)}_{u}(K)D_{\sigma}(P-K)\right.\nn
&\left.\hspace{2.5cm}-\ S^{(B)}_{u}(-K) D_{\pi^0}(P-K)\right.\nn
&\left.\hspace{2.5cm}-\ 2S^{(B)}_{d}(-K)D^{(B)}_{\pi^{\pm}}(P-K)\right], \label{sigma_u_intf}
\end{align}
and
\begin{align}
&\Sigma_d(B,P) = ig^2\int \frac{d^4K}{(2\pi)^4} \left[S^{(B)}_{d}(K)D_{\sigma}(P-K)\right.\nn
&\left.\hspace{2.5cm}-\ S^{(B)}_{d}(-K)D_{\pi^0}(P-K)\right.\nn
&\left.\hspace{2.5cm}-\ 2S^{(B)}_{u}(-K)D^{(B)}_{\pi^{\pm}}(P-K)\right]. \label{sigma_d_intf}
\end{align}
We make the change of variable $K^{\mu}\rightarrow -K^{\mu}$ in the second and third terms on the right-hand sides of both the Eqs.~\eqref{sigma_u_intf} and~\eqref{sigma_d_intf} and get 
\begin{align}
&\Sigma_u(B,P) = ig^2\int \frac{d^4K}{(2\pi)^4} \left[S^{(B)}_{u}(K)\bigg(D_{\sigma}(P-K)\right.\nn
&\left.\hspace{.5cm}-D_{\pi^0}(P+K)\bigg)-2S^{(B)}_{d}(K)D^{(B)}_{\pi^{\pm}}(P+K)\right], \nn
&\Sigma_d(B,P) = ig^2\int \frac{d^4K}{(2\pi)^4} \left[S^{(B)}_{d}(K)\bigg(D_{\sigma}(P-K)\right.\nn
&\left.\hspace{.5cm}-D_{\pi^0}(P+K)\bigg)-2S^{(B)}_{u}(K)D^{(B)}_{\pi^{\pm}}(P+K)\right]. \label{sigma_u,d}
\end{align}
The propagators of the $\pi^0$ and $\sigma$ fields are written in terms of Schwinger's proper-time parametrization as 
\begin{align}
D_{\pi^0}(K) &= -i\int\limits_{0}^{\infty} dt \,e^{it\(K^2-m^2_{\pi}+i\epsilon\)}, \\
D_{\sigma}(K) &= -i\int\limits_{0}^{\infty} dt \,e^{it\(K^2-m^2_{\sigma}+i\epsilon\)}.
\end{align}
Before diving into explicit calculations using Feynman diagrams (Fig.~\ref{fig:quark_self}), we note that the Dirac structure of the quark propagators along with its inverse (both bare and effective) and self-energies (which involve linear combinations of $\slashed{P}_{\shp}$, $\mathbb{1}$, $\slashed{P}_{\shp}\gamma^1\gamma^2$, $\gamma^1\gamma^2$ and $\slashed{P}_{\sperp}$) are the same, but the structure coefficients will be different. Following the aforementioned argument, we write down the structure of ${S^{*}_{\sF}}^{-1}$ as 
\begin{align}
{S^{*}_{\sF}}^{-1}(P) &= \mathcal{A}_{\sF}\slashed{P}_{\shp}+\mathcal{B}_{\sF}+\mathcal{C}_{\sF}\slashed{P}_{\shp}\gamma^1\gamma^2 \nn
&\hspace{2.5cm}+\mathcal{D}_{\sF}\gamma^1\gamma^2+\mathcal{E}_{\sF}\slashed{P}_{\sperp}. \label{gen_sigma}
\end{align}
Now we proceed to compute the momentum integrals in the expression of self-energy for a particular flavor $f$ which has three different channels, namely, the $\sigma$, $\pi^0$, and $\pi^{\pm}$ channels:  
\begin{align}
\Sigma_{[f\sigma]}(B,P) &= ig^2\int\dfrac{d^4K}{(2\pi)^4}S^{(B)}_{\sF}(K)D_{\sigma}(P-K),\\
\Sigma_{[f\pi^0]}(B,P) &= -ig^2\int\dfrac{d^4K}{(2\pi)^4}\nn
&\times\left[ S^{(B)}_{\sF}(K)D_{\pi^0}(P+K)\right],\\
\Sigma_{[f'\pi^{\pm}]} &= -2ig^2\int\dfrac{d^4K}{(2\pi)^4}\nn
&\times \left[S^{(B)}_{\sF}(K)D^{(B)}_{\pi^{\pm}}(P+K) \right].
\end{align} 
Now, referring back to Eq.~\eqref{gen_sigma}, we note that each of the structure coefficients appearing in it has
two parts: one comes from $S^{-1}_{\sF}$ and the other comes from $\Sigma_{f} = \Sigma_{[f\sigma]}+\Sigma_{[f\pi^0]}+\Sigma_{[f'\pi^{\pm}]}$, where
$f,f'=u,d$ with $f\neq f'$.
The structure coefficients coming from $\Sigma_{\sF}$, after performing the momentum integral, are calculated up to one-loop order. The results are quoted below in terms of two proper-time integrals:
\begin{widetext}  
	\begin{align}
	&\mathcal{A}^{(\Sigma)}_{\sF} = \frac{g^2}{(4\pi)^2}\int\limits_{0}^{\infty}\frac{dsdt\,t}{(s+t)^2}\left[\frac{\exp\(i\Phi_{\sF}^{(\sigma)}\)+\exp\(i\Phi_{\sF}^{(\pi^{0})}\)}{\widetilde{s}_{\sF}+t}+\frac{2 \exp\(i\Phi^{(\pi^{\pm})}_{\sF'}\)}{\(\tilde{s}_{\sF'}+\tilde{t}_e\)\cos\(|eB|t\)}\right], \label{AfS}\\
	&\mathcal{B}^{(\Sigma)}_{\sF} =M_{\sF} \frac{g^2}{(4\pi)^2}\int\limits_{0}^{\infty}\frac{dsdt}{s+t}\left[\frac{\exp\(i\Phi_{\sF}^{(\sigma)}\)-\exp\(i\Phi_{\sF}^{(\pi^{0})}\)}{\widetilde{s}_{\sF}+t}-\frac{2 \exp\(i\Phi^{(\pi^{\pm})}_{\sF'}\)}{\(\tilde{s}_{\sF'}+\tilde{t}_e\)\cos\(|eB|t\)}\right], \label{BfS}\\
	&\mathcal{C}^{(\Sigma)}_{\sF} = \frac{g^2}{(4\pi)^2}\int\limits_{0}^{\infty}\frac{dsdt\,t}{(s+t)^2}\left[\frac{\exp\(i\Phi_{\sF}^{(\sigma)}\)+\exp\(i\Phi_{\sF}^{(\pi^{0})}\)}{\(\widetilde{s}_{\sF}+t\)\cot\(|q_{\sF}B|s\)}\textsf{sgn}(q_{\sF}B)+\frac{2 \exp\(i\Phi^{(\pi^{\pm})}_{\sF'}\)\textsf{sgn}(q_{\sF'}B)}{\(\tilde{s}_{\sF'}+\tilde{t}_e\)\cos\(|eB|t\)\cot\(|q_{\sF'}B|s\)}\right], \label{CfS}\\
	&\mathcal{D}^{(\Sigma)}_{\sF} =M_{\sF} \frac{g^2}{(4\pi)^2}\int\limits_{0}^{\infty}\frac{dsdt}{(s+t)}\left[\frac{\exp\(i\Phi_{\sF}^{(\sigma)}\)-\exp\(i\Phi_{\sF}^{(\pi^{0})}\)}{\(\widetilde{s}_{\sF}+t\)\cot\(|q_{\sF}B|s\)}\textsf{sgn}(q_{\sF}B)-\frac{2 \exp\(i\Phi^{(\pi^{\pm})}_{\sF'}\)\textsf{sgn}(q_{\sF'}B)}{\(\tilde{s}_{\sF'}+\tilde{t}_e\)\cos\(|eB|t\)\cot\(|q_{\sF'}B|s\)}\right], \label{DfS}\\
	&\mathcal{E}^{(\Sigma)}_{\sF} = \frac{g^2}{(4\pi)^2}\int\limits_{0}^{\infty}\frac{dsdt}{(s+t)}\left[\frac{t \exp\(i\Phi_{\sF}^{(\sigma)}\)+t \exp\(i\Phi_{\sF}^{(\pi^{0})}\)}{\(\widetilde{s}_{\sF}+t\)^2\sec^2\(|q_{\sF}B|s\)}+\frac{2\widetilde{t}_{e} \exp\(i\Phi^{(\pi^{\pm})}_{\sF'}\)}{\(\tilde{s}_{\sF'}+\tilde{t}_e\)^2\cos\(|eB|t\)\cos^2\(|q_{\sF'}B|s\)}\right], \label{EfS}
	\end{align}
\end{widetext}
where we have defined
\begin{align}
&\wt{s}_{\sF} = \frac{\tan(|q_{\sF}B|s)}{|q_{\sF}B|}, \qquad \wt{t}_{e} = \frac{\tan(|eB|t)}{|eB|}, \label{tfsf}\\
&\Phi^{(\pi^{\pm})}_{\sF} = \frac{s\,t}{s+t}\,P_{\shp}^2+\frac{\wt{s}_{\sF}\, \wt{t}_{e}}{\wt{s}_{\sF}+\wt{t}_e}\,P_{\sperp}^2-sM_{\sF}^2-tm_{\pi}^2, \label{cPhipi_f}\\
&\Phi^{(\pi^{0})}_{\sF} = \frac{s\,t}{s+t}\,P_{\shp}^2+\frac{\wt{s}_{\sF}\, t}{\wt{s}_{\sF}+t}\,P_{\sperp}^2-sM_{\sF}^2-tm_{\pi}^2, \label{nPhipi_f} \\
&\Phi^{(\sigma)}_{\sF} = \frac{s\,t}{s+t}\,P_{\shp}^2+\frac{\wt{s}_{\sF}\, t}{\wt{s}_{\sF}+t}\,P_{\sperp}^2-sM_{\sF}^2-tm_{\sigma}^2. \label{nPhisigma_f}
\end{align} 
Note that there are two terms inside the square brackets in Eqs.~\eqref{AfS}-\eqref{EfS}: the first term comes from the diagrams where $\pi^0$ and $\sigma$ are inside the loop, and the second term comes from the diagrams where the charged pion is inside the loop. Also, in the second term the quark changes flavor from $f$ to $f^{\prime}$, with $f\neq f^{\prime}$, i.e., $u$ goes to $d$ [Fig.~\ref{fig:quark_self}(c)] and $d$ goes to $u$ [Fig.~\ref{fig:quark_self}(e)].\\
In the limit $B\rightarrow 0$, Eqs.~\eqref{AfS} -\eqref{EfS} reduce to   
\begin{align}
&\mathcal{A}_{\sF 0}^{(\Sigma)} = \frac{g^2}{(4\pi)^2}\int\limits_{0}^{\infty}\frac{dsdt\,t}{(s+t)^3}\left[\exp\(i\Phi^{(\sigma)}_{\sF 0}\) \right. \nn
&\hspace{2.cm}\left. +\exp\(i\Phi^{(\pi^0)}_{\sF 0}\) +2\exp\(i\Phi^{(\pi^{\pm})}_{\sF' 0}\)\right], \nn
&\mathcal{B}_{\sF 0}^{(\Sigma)} = \frac{g^2}{(4\pi)^2}\int\limits_{0}^{\infty}\frac{dsdt}{(s+t)^2}\left[\exp\(i\Phi^{(\sigma)}_{\sF 0}\) \right. \nn
&\hspace{2cm}\left. -\exp\(i\Phi^{(\pi^0)}_{\sF 0}\) -2\exp\(i\Phi^{(\pi^{\pm})}_{\sF' 0}\)\right], \nn
&\mathcal{C}_{\sF 0}^{(\Sigma)} = \mathcal{D}_{\sF 0}^{(\Sigma)}=0, \qquad	
\mathcal{E}_{\sF 0}^{(\Sigma)}=\mathcal{A}_{\sF 0}^{(\Sigma)}, \label{vac_struct_func}
\end{align}
where,
\begin{align}
&\Phi^{(\sigma)}_{\sF 0} = \frac{s \,t}{s+t}P^2-s\,M_{\sF}^2-t\,m^2_{\sigma}, \\
&\Phi^{(\pi^{\pm})}_{\sF 0} = \Phi^{(\pi^{0})}_{\sF 0} = \frac{s \,t}{s+t}P^2-s\,M_{\sF}^2-t\,m^2_{\pi}.
\end{align}
The ``$f 0$" in the subscript indicates that $f$ is the flavor of interest and the magnetic field is set to zero. 

The quark propagator in Eq.~\eqref{eq:full_sf} is written as
\begin{align}
S^{(B)}_{\sF}(P) &= \mathcal{A}^{(0)}_{\sF}\slashed{P}_{\shp}+\mathcal{B}^{(0)}_{\sF}+\mathcal{C}^{(0)}_{\sF}\slashed{P}_{\shp}\gamma^1\gamma^2\nn &\hspace{2cm}+\mathcal{D}^{(0)}_{\sF}\gamma^1\gamma^2+\mathcal{E}^{(0)}_{\sF}\slashed{P}_{\sperp},
\end{align}
where the coefficients are
\begin{align}
&\mathcal{A}^{(0)}_{\sF} = -i\int_{0}^{\infty} ds\,e^{i s\left\{P^2_{\shp}+\frac{\tan\(|q_fB|s\)}{|q_fB|s}P^2_{\sperp}-M^2_{\sF}\right\}}, \\
&\mathcal{B}^{(0)}_{\sF} = M_{\sF}\mathcal{A}^{(0)}_{\sF}, \\
&\mathcal{C}^{(0)}_{\sF} = -i\int_{0}^{\infty} ds\,e^{i s\left\{P^2_{\shp}+\frac{\tan\(|q_fB|s\)}{|q_fB|s}P^2_{\sperp}-M^2_{\sF}\right\}}\nn
&\hspace{3cm}\times\textsf{sgn}\({q_{\sF}B}\)\tan\(|q_{\sF}B|s\), \\
&\mathcal{D}^{(0)}_{\sF} = M_{\sF}\mathcal{C}^{(0)}_{\sF}, \\
&\mathcal{E}^{(0)}_{\sF} = -i\int_{0}^{\infty} ds\,e^{i s\left\{P^2_{\shp}+\frac{\tan\(|q_fB|s\)}{|q_fB|s}P^2_{\sperp}-M^2_{\sF}\right\}}\nn
&\hspace{4cm}\times \sec^2\(|q_{\sF}B|s\).
\end{align}
The inverse of the propagator $S^{-1}_{\sF}(P)$ has the same structure but different coefficients, and they are related to those of $S_{\sF}(p)$:
\begin{align}
&\mathcal{A}^{(0)}_{\sF,\text{in}}=\frac{\mathcal{A}^{(0)}_{\sF}}{\mathcal{F}_{\sF}}, \quad \mathcal{B}^{(0)}_{\sF,\text{in}}=-\frac{\mathcal{B}^{(0)}_{\sF}}{\mathcal{F}_{\sF}}, \quad
\mathcal{C}^{(0)}_{\sF,\text{in}}=-\frac{\mathcal{C}^{(0)}_{\sF}}{\mathcal{F}_{\sF}}, \quad \nn
&\hspace{1.4cm}\mathcal{D}^{(0)}_{\sF,\text{in}}=\frac{\mathcal{D}^{(0)}_{\sF}}{\mathcal{F}_{\sF}}, \quad
\mathcal{E}^{(0)}_{\sF,\text{in}}=\frac{\mathcal{E}^{(0)}_{\sF}}{\mathcal{F}_{\sF}},
\end{align}
where the ``in" subscript in the above set of equations refers to the structure coefficients corresponding to ${S^{(B)}_{\sF}}^{-1}$(P),
\begin{align}
\mathcal{F}_{\sF}&=\Big(P^2_{\shp}-M^2_{\sF}\Big)\left[\(\mathcal{A}^{(0)}_{\sF}\)^2+\(\mathcal{C}^{(0)}_{\sF}\)^2\right] \nn
&\hspace{3cm}+\(\mathcal{E}^{(0)}_{\sF}\)^2P^2_{\sperp}.
\end{align}
Since we are only interested in the magnetic field corrections, we subtract from  Eqs.~\eqref{AfS}-\eqref{EfS} their vacuum counterparts.
As a result of this, we have the following structure coefficients of ${S^{*}_{\sF}}^{-1}(P)$:
\begin{align}
\mathcal{A}_{\sF} &= \mathcal{A}^{(0)}_{\sF,\text{in}}-\mathcal{\hat{A}}^{(\Sigma)}_{\sF}, \\
\mathcal{B}_{\sF} &= \mathcal{B}^{(0)}_{\sF,\text{in}}-\mathcal{\hat{B}}^{(\Sigma)}_{\sF}, \\
\mathcal{C}_{\sF} &= \mathcal{C}^{(0)}_{\sF,\text{in}}-\mathcal{\hat{C}}^{(\Sigma)}_{\sF}, \\
\mathcal{D}_{\sF} &= \mathcal{D}^{(0)}_{\sF,\text{in}}-\mathcal{\hat{D}}^{(\Sigma)}_{\sF}, \\
\mathcal{E}_{\sF} &= \mathcal{E}^{(0)}_{\sF,\text{in}}-\mathcal{\hat{E}}^{(\Sigma)}_{\sF}.
\end{align} 
The hat symbol indicates a vacuum-subtracted structure coefficient of the constituent quark self-energy.
\\
The inversion process is shown explicitly in Appendix~\ref{sec:inversion}. 
\\

\subsection{Inversion of tensor structure in ${S^{*}_{\sF}}^{-1}$} \label{sec:inversion}
To achieve the inversion of Eq~\eqref{gen_sigma}, we first have to note the number of Dirac matrix structures involved in the expression. It has five tensor structures: $\slashed{P}_{\shp}$, $\slashed{P}_{\sperp}$, $\slashed{P}_{\shp}\gamma^1\gamma^2$, $\gamma^1\gamma^2$, and $\mathbb{1}$. The trick is to reduce the number of matrix structures as follows.

Suppose we have any matrix $\mathcal{M}$ whose inverse is desired. We multiply $\mathcal{M}$ by some matrix $\mathcal{R}$ of our choice and as a result we get some other matrix called $\mathcal{U}$: 
\begin{align}
\mathcal{M}\mathcal{R} = \mathcal{U}. \label{mru}
\end{align}\\
Note that we have to choose $\mathcal{R}$ in such a way that matrix structure of $\mathcal{U}$ is simple and $\mathcal{U}^{-1}$ can be found easily. So, from Eq.~\eqref{mru}, we get 
\begin{align}
\mathcal{M}^{-1}=\mathcal{R}\,\mathcal{U}^{-1}.
\end{align}
Thus, the essence of this technique lies in properly choosing $\mathcal{R}$.\\
 
In our case, we can identify ${S^{*}_{\sF}}^{-1}$ as $\mathcal{M}$ and choose $\mathcal{R}$ as  
\begin{align}
\mathcal{R}\equiv\mathcal{A}_{\sF}\slashed{P}_{\shp}+\mathcal{B}_{\sF}-\mathcal{C}_{\sF}\slashed{P}_{\shp}\gamma^1\gamma^2-\mathcal{D}_{\sF}\gamma^1\gamma^2+\mathcal{E}_{\sF}\slashed{P}_{\sperp}. \label{gen_sigma_p}
\end{align} 
According to Eq.~\eqref{mru}, we have  
\begin{align}
\mathcal{U}=\alpha+\beta_{\shp}\,\slashed{P}_{\shp}+\beta_{\sperp}\slashed{P}_{\sperp}-\wt{\beta}_{\sperp}\tilde{\slashed{P}}_{\sperp}, \label{eq:u}
\end{align}
where we have defined 
\begin{align}
\tilde{P}^{\mu} \equiv \frac{1}{B}F^{\mu\nu}P_{\nu} = (0,p^2,-p^1,0)
\end{align}
 with $F^{\mu\nu}$ denoting Maxwell's electromagnetic field-strength tensor ($F^{21}=-F^{12}=B$ and zero for others) in our background field configuration and 
\begin{align}
\alpha &\equiv \(\mathcal{A}_{\sF}^2+\mathcal{C}_{\sF}^2\)P^{2}_{\shp}+\mathcal{E}_{\sF}^2P_{\sperp}^2+\mathcal{B}_{\sF}^2+\mathcal{D}_{\sF}^2, \nn
\beta_{\shp} &\equiv 2\(\mathcal{A}_{\sF}\mathcal{B}_{\sF}+\mathcal{C}_{\sF}\mathcal{D}_{\sF}\), \nn
\beta_{\sperp} &\equiv 2\mathcal{B}_{\sF}\mathcal{E}_{\sF}, \nn
\tilde{\beta}_{\sperp} &\equiv 2\mathcal{D}_{\sF}\mathcal{E}_{\sF}.
\end{align}
The inversion of $\mathcal{U}$ is easy to perform and it is given as 
\begin{align}
\mathcal{U}^{-1} = \frac{\beta_{\shp}\,\slashed{P}_{\shp}+\beta_{\sperp}\slashed{P}_{\sperp}-\tilde{\beta}_{\sperp}\tilde{\slashed{P}}_{\sperp}-\alpha}{\beta_{\shp}^2P^2_{\shp}+\(\beta^2_{\sperp}+\tilde{\beta}_{\sperp}^2\)P^2_{\sperp}-\alpha^2}.
\end{align}
Finally, we get $S^{*}_{\sF}$ as 
\begin{widetext}
	\begin{align}
		S^{*}_{\sF}(P) = \(\mathcal{A}_{\sF}\slashed{P}_{\shp}+\mathcal{B}_{\sF}-\mathcal{C}_{\sF}\slashed{P}_{\shp}\gamma^1\gamma^2-\mathcal{D}_{\sF}\gamma^1\gamma^2+\mathcal{E}_{\sF}\slashed{P}_{\sperp}\)\frac{\beta_{\shp}\,\slashed{P}_{\shp}+\beta_{\sperp}\slashed{P}_{\sperp}-\tilde{\beta}_{\sperp}\tilde{\slashed{P}}_{\sperp}-\alpha}{\beta_{\shp}^2P^2_{\shp}+\(\beta^2{\sperp}+\tilde{\beta}_{\sperp}^2\)P^2_{\sperp}-\alpha^2}.
	\label{q_prop_eff}
	\end{align}
\end{widetext}
The one-loop effective quark mass $M_{f, \rm eff}$ in the presence of a magnetic field is obtained by solving for the denominator of Eq.~\eqref{q_prop_eff}, i.e.,
\begin{equation}
\beta_{\shp}^2P^2_{\shp}+\(\beta^2{\sperp}+\tilde{\beta}_{\sperp}^2\)P^2_{\sperp}-\alpha^2=0
\end{equation}
 for $p_0$ in the static limit i.e., $\bf{p}\rightarrow 0$.
 

\section{Vacuum regularization}
In this appendix, we discuss the regularization of one particular diagram, namely, the vacuum contribution of the meson loop in Eq.~\eqref{eq:pi_pieB02} using the method of dimensional regularization. In $d=4-2\epsilon$ dimensions, Eq.~\eqref{eq:pi_pieB02} becomes
\begin{align}
\Pi_{\pi^{\pm}}^{\rm vac} = \frac{\lambda}{4}\int\frac{d^dK}{(2\pi)^d}\frac{i}{K^2-m^2_{\pi}+i\varepsilon}. \label{eq:pi_pieB_d}
\end{align}
In Euclidean space,
\begin{align}
\Pi_{\pi^{\pm}}^{\rm vac} = \frac{\lambda}{4}\int\frac{d^dK_E}{(2\pi)^d}\frac{1}{K_E^2+m_{\pi}^2}. \label{eq:pi_pieB_d_Euc}
\end{align}
It is very standard to carry out the integral and the result is
\begin{align}
\Pi_{\pi^{\pm}}^{\rm vac} = \frac{\lambda m_{\pi}^2}{64 \pi^2}\(\frac{m^2_{\pi}}{4\pi \Lambda^2}\)^{-\epsilon}\Gamma\(\epsilon - 1\) \label{eq:pi_pieB_d_euc_exact}
\end{align}
Here $\Lambda$ is the renormalization scale, and $\Gamma$ is the gamma function defined as 
\begin{align}
\Gamma(z) = \int\limits_{0}^{\infty} dx\, x^{z-1}e^{-x},
\end{align} 
which is analytic in the complex $z$ plane except for simple poles at $z=0,-1,-2,-3,...$\\
The right hand side of Eq.~\eqref{eq:pi_pieB_d_euc_exact} is divergent at $\epsilon= 0$ or $d=4$. We can isolate the divergent piece by taking the limit $\epsilon\rightarrow 0$ and keeping the leading-order term. Thus,
\begin{align}
\Pi_{\pi^{\pm}}^{\rm vac} = - \frac{\lambda m_{\pi}^2}{64 \pi^2} \left[\frac{1}{\epsilon}+1-\gamma_E-\ln\(\frac{m_{\pi}^2}{4\pi\Lambda^2}\)\right], 
\end{align}\\
where $\gamma_E$ is the Euler-Mascheroni constant.\\

Finally, using the $\overline{\rm{MS}}$ scheme with the ultraviolet scale $\Lambda$, the vacuum part of the one-loop pure charged meson contribution is written as
\begin{align}
\Pi_{\pi^{\pm}}^{\rm vac} = - \frac{\lambda m_{\pi}^2}{64 \pi^2} \left[\ln\(\frac{\Lambda^2}{m_{\pi}^2}\)\right].
\end{align}

In weak field limit, $\mathcal{O}(eB)^2$ term of the charged meson loop is
\begin{align}
\Pi_{\pi^{\pm}}^{ (eB)^2} = -\frac{\lambda}{4}\int\frac{d^dK}{(2\pi)^d}\frac{i(eB)^2}{\(K^2-m^2_{\pi}\)^2}, \label{eq:pi_pieB2}
\end{align}
which is UV finite and can be evaluated at $\epsilon\rightarrow 0$ as
\begin{align}
\Pi_{\pi^{\pm}}^{ (eB)^2} =  -\frac{\lambda }{384 \pi^2} \frac{(eB)^2}{m_{\pi}^2}.
\end{align}
All of the higher-order terms  as well as the general magnetic field expression for the charged pion self-energy are also UV finite. This is true for all of the diagrams. Combining all of the renormalized vacuum parts of the self-energy, one can study the pion mass in vacuum. In the present calculation, we are interested in the magnetic field’s effect on the vacuum pion mass, and we take it to be $\sim 140$ MeV.


\end{document}